\pdfoutput=1
\documentclass[conference,compsoc]{IEEEtran}

\usepackage{amsmath,amssymb}
\usepackage{graphicx}
\usepackage{subfig}
\usepackage{booktabs,tabularx,array,threeparttable,diagbox}
\usepackage{dblfloatfix}
\usepackage{enumitem}
\usepackage{pifont,marvosym,xspace,url}
\usepackage[table]{xcolor}
\usepackage[hidelinks]{hyperref}

\ifCLASSOPTIONcompsoc
  \usepackage[nocompress]{cite}
\else
  \usepackage{cite}
\fi

\providecommand{\safeincludegraphics}[2][]{%
  \IfFileExists{#2}{%
    \includegraphics[#1]{#2}%
  }{%
    \fbox{\parbox[c][0.75in][c]{0.95\linewidth}{%
      \centering\scriptsize\ttfamily\detokenize{#2}}}%
  }%
}

\definecolor{scoreblue}{RGB}{0,82,183}
\newcommand{\ScoreCell}[1]{%
  \cellcolor{scoreblue!#1!white}\rule{0pt}{1.55ex}{\fontsize{8}{8}\selectfont #1}%
}

\begin{document}
\nocite{IEEEexample:BSTcontrol} 

\title{Your Camera Can Hear You: Side-Channel Acoustic Eavesdropping \\via Android Camera2 Metadata}

\author{
    \IEEEauthorblockN{
        Guangyuan Ji\IEEEauthorrefmark{1}, 
        Wenjing Wang\IEEEauthorrefmark{2}, 
        and Bingsheng Zhang\IEEEauthorrefmark{1}\textsuperscript{,\Letter}
    }
    \vspace{1.5mm}
    
    \IEEEauthorblockA{\IEEEauthorrefmark{1}The State Key Laboratory of Blockchain and Data Security, Zhejiang University, Zhejiang, China}
    \IEEEauthorblockA{\IEEEauthorrefmark{2}Electric Power Research Institute of Guangdong Power Grid, Guangdong, China}
    \IEEEauthorblockA{Email: \{guangyuanji, bingsheng\}@zju.edu.cn, wenjing\_wang@whu.edu.cn}

}

\newcommand{\sysname}{\textsc{CamEar}\xspace}

\maketitle

\begin{abstract}

To prevent motion sensors from acting as clandestine microphones, modern Android enforces strict defensive boundaries around \texttt{SensorManager}. Since Android~12, gyroscope and accelerometer rates are capped at 200~Hz by default, requesting higher rates requires the heavily scrutinized \texttt{HIGH\_SAMPLING\_RATE\_SENSORS} permission, and toggling off the system-wide microphone privacy switch throttles motion sensors back to 200~Hz even when that permission is granted. 
In this paper, we show that Android's Camera2 subsystem bypasses these speech-privacy defenses. During an active Camera2 session, optical and electronic image stabilization pipelines in the Camera HAL record high-rate gyroscope measurements and export them to user space via vendor-defined \texttt{CaptureResult} metadata. When speech plays through the phone's loudspeaker, chassis vibrations couple into this stabilization path. Across 50 smartphones (2020--2026) from 10 vendors, 37 expose gyroscope metadata via Camera2, and 26 exceed the 200~Hz safety limit---including six recent flagship models streaming at 800--1000~Hz, double the 400--500~Hz hardware ceiling of \texttt{SensorManager}. Crucially, this side channel requires neither \texttt{RECORD\_AUDIO} nor \texttt{HIGH\_SAMPLING\_RATE\_SENSORS} and defeats Android's system-wide privacy switches. Namely, turning off the microphone toggle fails to throttle Camera2 metadata, and turning off the camera toggle blanks only video frames while 800--1000~Hz metadata continues streaming uninterrupted.

To turn this vulnerability into a practical, zero-profiling attack that generalizes to unseen phone models, we introduce \emph{Source-to-Leakage Learning} (S2L). Unlike prior attacks that train on real sensor traces---where speech content is inextricably entangled with a specific phone's mechanical resonances and sampling rate---S2L decouples speech content from hardware acoustics using a physics-constrained forward simulator. Trained \emph{exclusively} on simulated leakage without any target-phone traces, our universal recognizer \textsc{CamEar} achieves 52.1\% mean word accuracy (up to 61.8\%) and 74.3\% character accuracy (up to 80.7\%) across the six 800--1000~Hz phones. In contrast, state-of-the-art ISpyU models fine-tuned on real target-phone data drop to 25.9\% word accuracy on unseen phones; \textsc{CamEar} improves cross-device accuracy by 101.4\% and even outperforms ISpyU's same-device models on every tested phone.

\end{abstract}

\section{Introduction}
\label{sec:intro}
Over the past decade, numerous studies have shown that smartphone motion sensors can be exploited as acoustic side channels; when speech plays through a phone's built-in loudspeaker, mechanical vibrations travel through the chassis and modulate the onboard gyroscope and accelerometer~\cite{michalevsky2014gyrophone, anand2021spearphone, ba2020learning, hu2022accear, zhang2023spy}. Recognizing this threat, Google built a multi-layered defense around Android's \texttt{SensorManager}. Since Android 12, motion sensor delivery has been capped at 200~Hz by default~\cite{android_sensor_rate_limiting}, a rate at which severe spectral aliasing destroys the harmonic structure needed for continuous speech recognition. Accessing higher rates (400--500~Hz) requires declaring the \texttt{HIGH\_SAMPLING\_RATE\_SENSORS} manifest permission, which triggers Google Play Store review warnings~\cite{google_high_sampling_lint}. 
Furthermore, Android 12 introduced system-wide microphone and camera privacy toggles (``kill switches''): when a user turns off microphone access, Android forcibly throttles \texttt{SensorManager} back to $\le$200~Hz across all apps, even for those holding \texttt{HIGH\_SAMPLING\_RATE\_SENSORS}~\cite{android_sensor_rate_limiting}. Together, default rate caps, store-level permission scrutiny, and system-wide privacy toggles were widely believed to have neutralized high-rate motion-sensor eavesdropping on modern Android devices.

\smallskip
\noindent\textbf{The Camera2 Metadata Bypass.}
In this work, we demonstrate that Android's Camera2 subsystem unwittingly exposes an unthrottled side channel that bypasses all of these protections. To compensate for hand shake during video capture, optical and electronic image stabilization (OIS/EIS) pipelines in the Camera Hardware Abstraction Layer (HAL) sample the physical gyroscope at high rates. More specifically, the Camera HAL packages these batched, timestamped 3-axis gyroscope readings into vendor-defined \texttt{CaptureResult} metadata fields, which Android's \texttt{CameraMetadataNative} framework passes directly to user-space applications~\cite{android_capture_result}. When loudspeaker audio excites the phone chassis, these vibrations modulate the stabilization gyroscope stream. As a result, during an active Camera2 session, an untrusted app or an embedded third-party SDK (e.g., vision, AR, or RTC libraries) in a benign camera host can harvest high-rate, speech-modulated gyroscope traces through \texttt{CAMERA} alone, without \texttt{RECORD\_AUDIO} or \texttt{HIGH\_SAMPLING\_RATE\_SENSORS}, even when system microphone and camera toggles are off.

To understand the real-world scope of this flaw, we conduct an empirical study of 50 Android smartphones released between 2020 and 2026 across 10 OEMs and five SoC families. Our measurements reveal a pervasive architectural blind spot caused by inconsistent cross-API mediation:
\begin{itemize}[leftmargin=0pt, itemindent=0pt, nosep]

\item \textbf{Widespread high-rate exposure:} Out of 50 phones, 37 expose gyroscope metadata via Camera2, and 26 exceed Android's 200~Hz safety cap. Remarkably, six recent models (2025--2026) stream at \textbf{800--1000~Hz}, double the 400--500~Hz ceiling of standalone \texttt{SensorManager} even with \texttt{HIGH\_SAMPLING\_RATE\_SENSORS}. At 800--1000~Hz, the 400--500~Hz Nyquist band resolves unaliased pitch harmonics and nearly quadruples spectral capacity relative to the 200~Hz cap.
\item \textbf{Complete bypass of system-wide privacy toggles:} Cross-API provenance analysis confirms that Camera2 metadata and \texttt{SensorManager} read from the \emph{exact same} upstream micro-electro-mechanical systems (MEMS) gyroscope sequence ($R^2 \ge 0.9993$). Yet, Android's privacy switches mediate only the primary sensor/image paths: turning off the system-wide microphone toggle throttles \texttt{SensorManager} to 200~Hz, and turning off the camera toggle replaces video frames with blank buffers, \emph{but 800--1000~Hz Camera2 metadata continues streaming unabated under both toggles on supporting phones}. Users disabling both privacy toggles for a camera session remain exposed to loudspeaker eavesdropping.
\item \textbf{Hardware-level rate spillover:} Because the underlying sensor hub lacks per-client rate isolation, opening a foreground Camera2 session switches the shared gyroscope into an 800--1000~Hz state, boosting concurrent \texttt{SensorManager} listeners from 200~Hz to 400--500~Hz without \texttt{HIGH\_SAMPLING\_RATE\_SENSORS} and to 800--1000~Hz with the permission.
\end{itemize}

\smallskip
\noindent\textbf{The Cross-Device Portability Bottleneck.}
Turning this unthrottled 800--1000~Hz channel into a practical threat confronts a long-standing fundamental challenge in learning-based side-channel analysis: \emph{zero-profiling cross-device portability}~\cite{bhasin2020mind, picek2023sok, karayalcin2026sok}. In practice, an adversary cannot access a victim's handset in advance to collect tens of hours of paired audio-vibration training data, yet transferring a side-channel recognizer to an unseen handset typically causes a severe accuracy drop~\cite{bhasin2020mind}. Each handset exhibits a distinct physical transfer path $\mathcal{H}_d$ governed by chassis geometry, MEMS resonance modes, damping ratios, and exact sampling rates (e.g., 799~Hz vs.\ 998~Hz, which fold upper harmonics to different baseband frequencies). Existing attacks rely on measurement-driven training: they either fine-tune on target-device traces~\cite{zhang2023spy, cao2021cross, yu2021cross} or pool data from a handful of training devices~\cite{bhasin2020mind, das2019x, su2021towards}. Because every recorded trace $\ell_d = \mathcal{H}_d(s) + \eta$ (with measurement noise $\eta$) permanently binds the spoken content $s$ with the specific hardware response $\mathcal{H}_d$ of the collection device, models trained on real traces inevitably overfit to the training hardware.

\smallskip
\noindent\textbf{Our Solution: Source-to-Leakage Learning.}
To achieve open-vocabulary speech recognition on unseen phones with \emph{zero} target-device profiling, we revisit how side-channel training data is constructed. Our key insight is that while speech content $s$ and hardware transfer operator $\mathcal{H}_d$ are entangled after sensor measurement, they are physically independent factors prior to measurement. Building on this insight, we propose the \textbf{Source-to-Leakage Learning (S2L)} paradigm. Instead of collecting real sensor traces from target hardware, S2L models the physical acquisition pipeline through a parameterized, physics-constrained forward operator family $\{\widetilde{\mathcal{H}}_\theta : \theta \sim p_\Theta\}$ capturing multi-mode chassis--MEMS structural resonances, cross-axis readout projection, rate-adaptive sub-Nyquist metadata sampling ($\approx$760--1100~Hz), and quantization noise. During training, S2L independently draws labeled public speech $(s, y) \sim p_S$ and channel parameters $\theta \sim p_\Theta$ to synthesize $(\widetilde{\mathcal{H}}_\theta(s), y)$ on the fly. By decoupling linguistic scale from hardware diversity, S2L trains a single end-to-end recognizer, \sysname, which generalizes across unseen phone models without ever seeing a single real gyroscope trace during training.

Evaluated across the six 800--1000~Hz phones (4{,}896 measured test traces), our universal \sysname recognizer trained purely on simulated leakage achieves \textbf{52.1\% mean word accuracy} (WAcc, up to \textbf{61.8\%}) and \textbf{74.3\% mean character accuracy} (CAcc, up to \textbf{80.7\%}), rising to \textbf{60.1\% mean WAcc} with text-only large language model (LLM) refinement. In contrast, state-of-the-art ISpyU models fine-tuned on 24~hours of real Camera2 traces from a profiling phone achieve only \textbf{25.9\% cross-device mean WAcc} across the other five phones. \sysname thus more than doubles cross-device accuracy (+101.4\%) without target profiling, and even outperforms ISpyU's 24-hour \emph{same-device} fine-tuned models (36.6\% mean WAcc) on every evaluated phone. Finally, capping Camera2 metadata reduces WAcc from 52.1\% to 32.4\% at 400--499~Hz and to 2.3\% under Android's 200~Hz limit, providing a concrete remediation path that we have disclosed to the affected vendors.

\smallskip
\noindent\textbf{Our Contributions.} The main contributions of this work can be summarized as follows:
\begin{itemize}[leftmargin=0pt, itemindent=0pt, nosep]
\item \textbf{Uncovering the Camera2 Acoustic Side Channel.} We identify a frame-independent acoustic side channel in Camera2 \texttt{CaptureResult} metadata across 50 phones (with 26 exceeding the 200~Hz cap and six streaming at 800--1000~Hz) that bypasses Android's motion-sensor rate caps, manifest permissions, and system-wide microphone/camera privacy toggles.
\item \textbf{Source-to-Leakage Learning (S2L) Paradigm.} We formulate S2L to overcome the cross-device portability bottleneck in physical side channels. By decoupling large-scale linguistic supervision from a physics-constrained forward model of chassis--MEMS resonances and sub-Nyquist aliasing, S2L eliminates the need for target-hardware data collection.
\item \textbf{Zero-Profiling Cross-Device Speech Eavesdropping.} We implement \sysname and demonstrate that a single simulation-trained recognizer improves cross-device WAcc by \textbf{101.4\%} over 24-hour fine-tuned ISpyU baselines (from 25.9\% to 52.1\%) and surpasses same-device fine-tuning on every evaluated phone.
\end{itemize}

\section{Background and Related Work}
\label{sec:background}

\subsection{Camera Stabilization and Capture Metadata}
OIS and EIS systems compensate for handset and camera-module motion during video capture. In Android's Camera2 pipeline, applications submit \texttt{CaptureRequest}s and receive output image buffers together with \texttt{CaptureResult} metadata describing the sensor, lens, and processing state~\cite{android_capture_result}. Alongside standard Android Open Source Project (AOSP) fields, SoC and handset vendors define custom metadata keys in the Camera HAL to pass batched, timestamped three-axis gyroscope samples for stabilization, calibration, and diagnostics. Because AOSP's \texttt{CameraMetadataNative} marshals all populated vendor keys to user space without filtering~\cite{android_capture_result_keys,android_vendor_keys_aosp}, any app holding \texttt{CAMERA} can read these high-rate motion sequences from each \texttt{CaptureResult}.

\subsection{Motion-Sensor Rate Controls and Permission Boundaries}
Android governs visual, audio, and inertial interfaces through distinct permission models. Opening a Camera2 session requires the runtime permission \texttt{CAMERA}, and capturing audio requires \texttt{RECORD\_AUDIO}. Public motion sensors exposed through \texttt{SensorManager}, by contrast, are governed by sampling-rate tiers rather than a runtime prompt~\cite{android_motion_sensors}. To support ordinary UI and gaming motion while mitigating acoustic eavesdropping, Android~12 caps \texttt{SensorManager} gyroscope and accelerometer delivery at 200~Hz by default~\cite{android_sensor_rate_limiting}. Requesting higher rates (typically 400--500~Hz) requires declaring the manifest permission \texttt{HIGH\_SAMPLING\_RATE\_SENSORS}, which triggers official Android lint warnings and potential Google Play review scrutiny~\cite{google_high_sampling_lint}; in practice, fewer than 4\% of the applications that used the fastest sensor rate on Android~10 declared this permission on Android~12~\cite{long2025armour}. At the hardware layer, Android multiplexes a single physical inertial sensor across concurrent clients. Prior work showed that within \texttt{SensorManager}, the highest sampling rate requested by one client can also be delivered to concurrent lower-rate clients~\cite{shepherd2023side,long2025armour}.

Android~12 also introduced system-wide microphone and camera privacy toggles in the system UI~\cite{android_privacy_controls}. Disabling microphone access blocks audio capture and enforces the 200~Hz cap on \texttt{SensorManager} across all applications, even when an application holds \texttt{HIGH\_SAMPLING\_RATE\_SENSORS}~\cite{android_sensor_rate_limiting}. Disabling camera access causes Camera2 to return blank image buffers so that camera-dependent applications can remain open without capturing visual scenes~\cite{android_privacy_controls}. 

\subsection{Cross-Device Portability in Physical Side Channels}
\label{sec:background_portability}
Cross-device portability, defined as the ability of one recognition pipeline to remain effective across physical instances without instance-specific leakage profiling or adaptation, is a central challenge in learning-based side-channel analysis~\cite{bhasin2020mind,picek2023sok,karayalcin2026sok}. Same-device evaluation can overestimate attack accuracy by an order of magnitude~\cite{bhasin2020mind}, because every measured trace convolves the target signal with the physical transfer response of the specific test hardware. This hardware dependence spans chips and probes in power and electromagnetic (EM) analysis~\cite{bhasin2020mind,das2019x,danial2021x}, optical paths in visual eavesdropping~\cite{davis2014visual,nassi2022lamphone}, propagation geometry in radio-frequency sensing~\cite{basak2022mmspy, wang2024vibspeech}, and chassis--MEMS dynamics in motion sensors~\cite{michalevsky2014gyrophone,ba2020learning}.

To generalize across hardware, existing measurement-based attacks pool traces from multiple training devices~\cite{bhasin2020mind,das2019x,danial2021x}, learn device-invariant features~\cite{su2021towards,cao2022pa}, or fine-tune on target-device data~\cite{cao2021cross,yu2021cross}. However, traces recorded on a small set of phones cover only discrete hardware instances and fixed sampling rates, failing when an unseen handset shifts modal resonances or aliasing fold frequencies. Section~\ref{sec:systemdesign} addresses this limitation by modeling forward chassis--MEMS and sampling physics directly over a continuous prior.

\subsection{Smartphone Acoustic Side Channels}
\noindent\textbf{Camera images and camera-related emissions.}
Optical-acoustic attacks recover sound from pixel distortion and vibrations of objects in a scene~\cite{akutsu2013extract,davis2014visual,zhang2016note,zhang2017efficient}. Side Eye shows that structure-borne sound affects rolling-shutter imaging and movable camera components~\cite{long2023side}. OISMic measures external EM leakage caused by sound-induced OIS vibration~\cite{shao2024oismic}. \sysname operates purely on numeric \texttt{CaptureResult} gyroscope metadata inside user space, requiring no scene visibility or external probes.

\noindent\textbf{Public motion-sensor streams.}
Smartphone accelerometers and gyroscopes capture speech-induced vibration through shared-surface and chassis-coupled paths exposed by public motion-sensor APIs~\cite{michalevsky2014gyrophone,anand2018speechless,anand2021spearphone,zhang2015accelword,liang2022escalated,gao2022inertiear,hu2022accear,sun2023stealthyimu,yao2024watch,najeeb2025glitch,han2017pitchln,cao2023live}. At sampling rates near 200~Hz, prior attacks primarily perform closed-set digit or hotword classification using coarse whole-word spectral envelopes~\cite{michalevsky2014gyrophone,ba2020learning,gao2022inertiear}. At higher rates, prior continuous-speech and audio-reconstruction attacks~\cite{hu2022accear,liang2022escalated,zhang2023spy} all assume prior access to the target device to collect paired speech--sensor traces, which is impractical in the wild. Even ISpyU~\cite{zhang2023spy}, despite fixed-decimation synthetic pretraining, still requires tens of hours of target-device fine-tuning traces. By contrast, \sysname exploits unthrottled 800--1000~Hz Camera2 metadata for zero-profiling cross-device speech transcription.

\noindent\textbf{Broader unintended sensing.}
Research on sensor out-of-band vulnerabilities studies how physical transduction turns environmental stimuli, such as acoustic waves or EM interference, into unintended signals within non-acoustic sensing and I/O components~\cite{zheng2026phyfuzz, xiao2025sok}. Related attacks exploit circuit EM emanations~\cite{long2024eye,chen2024eavesdropping}, optical mouse sensors~\cite{fakih2025invisible}, or active carrier injection across nonlinear circuits (with target-model frequency profiling)~\cite{yan2026injected} to recover analog signals. Complementing hardware-level transduction studies, our work shows that even when an inertial sensor operates as intended for camera stabilization, exporting its unthrottled telemetry across an unmediated software interface (\texttt{CaptureResult}) creates a high-rate acoustic side channel that bypasses platform sensor controls.

\section{Gyro-Related Camera2 Metadata in the Wild}
\label{sec:measurement}

We survey gyro-related Camera2 metadata across 50 phones, examine how the 800--1000~Hz metadata streams preserve loudspeaker-excited acoustic structure across sampling-rate tiers, relate them to public gyroscope events, and evaluate their behavior across Android's sensor-control and application boundaries.

\subsection{Measurement Methodology}
\label{sec:measurement_method}

\noindent\textbf{Device set.}
We test 50 phones released between 2020 and 2026, spanning 10 vendors, 14 marketed Android product lines, and five SoC families. For each phone, we record its model, SoC family, exposed vendor-defined \texttt{CaptureResult} key, and nominal metadata rate tier; Appendix~\ref{sec:appendix_device_matrix} (Table~\ref{tab:appendix_device_matrix}) reports the phone-level matrix.

\noindent\textbf{Camera2 measurement application.}
Our measurement application targets Android API level~35 and requests only the \path{CAMERA} permission. It (1)~opens a default rear-camera preview (where handset OIS/EIS resides), (2)~enumerates visible \path{CaptureResult} keys via \path{getAvailableCaptureResultKeys()}~\cite{android_capture_result_keys,android_vendor_keys_aosp}, and (3)~decodes numeric and packed-byte vendor payloads. Because Camera2 delivers \path{CaptureResult} callbacks at the preview frame rate ($\approx 30$~frames per second, fps), high-rate HAL implementations batch multiple timestamped three-axis gyroscope samples into each per-frame metadata payload. We identify gyro-related candidates from vendor-defined key names and payload structures, retaining fields whose batched samples form continuous timelines throughout the preview. Using the embedded hardware timestamps, the application restores within-batch sample order and estimates the effective metadata sampling rate (Appendix~\ref{sec:appendix_rate_estimation}).

\noindent\textbf{Cross-API diagnostics.}
To compare Camera2 metadata against public gyroscope delivery, a diagnostic build simultaneously records the selected Camera2 metadata stream and \texttt{SensorManager} \texttt{TYPE\_GYROSCOPE} events, tested both with and without \texttt{HIGH\_SAMPLING\_RATE\_SENSORS}. Privacy-control tests independently disable the system-wide microphone or camera toggle both during an active Camera2 session and when opening a new session. For cross-application tests, a concurrent non-camera application (via a foreground service) records \texttt{SensorManager} \texttt{TYPE\_GYROSCOPE} events while a separate foreground camera application opens or closes a Camera2 session.

\subsection{Exposure Landscape}
\label{sec:exposure_landscape}

Of the 50 tested phones, 37 expose a continuous gyro-related Camera2 metadata stream: eight provide only a frame-bound field near 30~Hz, three stream at 90--100~Hz, 20 at 400--500~Hz, and six at 800--1000~Hz. Notably, all six 800--1000~Hz handsets are 2025--2026 releases (Table~\ref{tab:appendix_device_matrix}), indicating that higher-rate stabilization metadata is increasingly adopted in modern camera pipelines. Sections~\ref{sec:cross_api_provenance}--\ref{sec:policy_boundaries} examine these six models in detail, as they open a new rate tier beyond \texttt{SensorManager}'s 400--500~Hz ceiling.

Although Android limits gyroscope events delivered through \path{SensorManager.registerListener()} to 200~Hz by default unless the application declares \texttt{HIGH\_SAMPLING\_RATE\_SENSORS}~\cite{android_sensor_rate_limiting}, our measurement application receives gyro-related Camera2 metadata streams at 400--1000~Hz on 26 phones. These 26 phones use a small set of SoC- and OEM-wide key names and payload structures (Table~\ref{tab:appendix_device_matrix}), such as \path{com.mediatek.3afeature.gyrodata}, \path{vivo.parameter.captureGyroData}, and \path{xiaomi.mivi.ellc.gyro}. However, phones sharing the same SoC generation or vendor-tag namespace expose rates spanning 400--1000~Hz (for example, Dimensity~9500 models operate at 400~Hz, 500~Hz, 800~Hz, 960~Hz, or 1000~Hz), showing that the exposed rate depends on per-model HAL configuration rather than SoC family alone.

\noindent\textbf{Finding 1: An application holding only \texttt{CAMERA} obtains 400--1000~Hz gyro-related metadata on 26 of 50 phones without declaring \texttt{HIGH\_SAMPLING\_RATE\_SENSORS} or registering a \texttt{SensorManager} gyroscope listener.}

\subsection{Cross-API Provenance}
\label{sec:cross_api_provenance}

We hypothesize that Camera2 metadata and \texttt{SensorManager} gyroscope events tap the exact same upstream MEMS motion sequence, and verify this hypothesis across 12 synchronized Camera2--\texttt{SensorManager} sessions (two per phone). For each session, we estimate a constant inter-API clock offset and fit a one-dimensional affine mapping per axis on the first half of the recording, holding both fixed for evaluation on the second half. Across the held-out intervals on all six phones, every \texttt{SensorManager} timestamp matches a unique Camera2 metadata timestamp within 100~ns after offset correction, forming a one-to-one, lower-rate subset of the Camera2 sequence. At matched timestamps, the held-out affine mappings achieve $R^2 \ge 0.9993$ across all axes, sessions, and phones, reflecting constant unit scaling between raw HAL metadata values and the angular velocity in $\mathrm{rad/s}$ reported by \texttt{SensorManager}.

These timestamp alignments and affine relationships confirm that both APIs read from the same upstream MEMS gyroscope sequence: individual samples at matched timestamps share an identical sensor noise floor, while Camera2 exposes a $2\times$ to $5\times$ denser temporal view of that sequence than standalone \texttt{SensorManager}.

\noindent\textbf{Finding 2: Camera2 and \texttt{SensorManager} expose timestamp-aligned, different-rate views derived from a shared upstream motion sequence.}

\subsection{Acoustic Response and Aliasing}
\label{sec:acoustic_modulation}

\begin{figure}[t]
    \centering
    \includegraphics[width=\columnwidth]{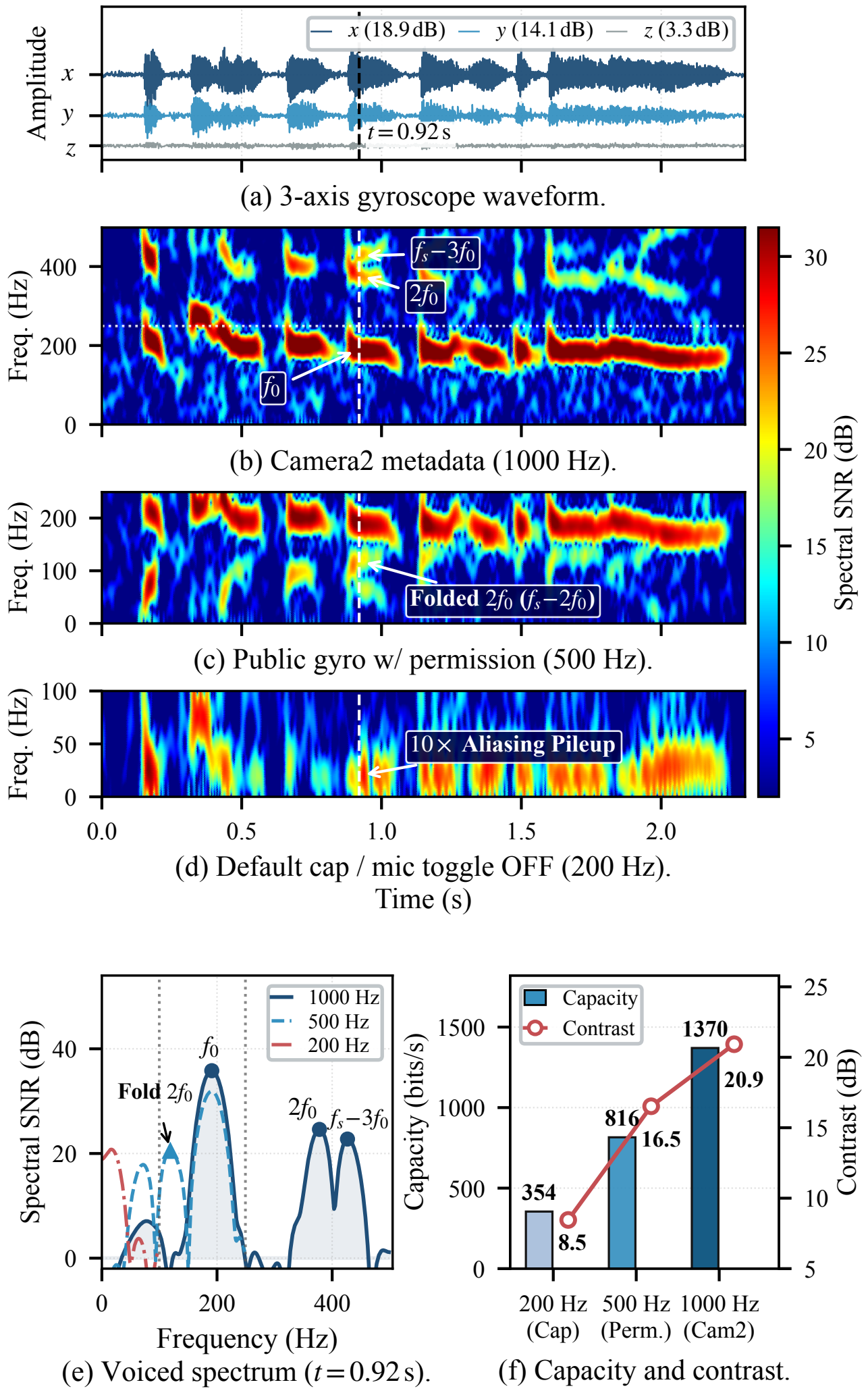}
\caption{Physical and spectral characterization of speech-modulated gyroscope traces across sampling-rate tiers (vivo X300s, utterance \texttt{84-121123-0022}).}
    \label{fig:multirate_physics}
\end{figure}

Recovering continuous speech from gyroscope metadata requires two physical conditions: the structural--MEMS path must couple speech harmonics above the sensor noise floor, and the metadata rate must keep short-time harmonic trajectories separable despite sub-Nyquist folding. Figure~\ref{fig:multirate_physics} characterizes both factors on the vivo X300s ($f_s \approx 998$\,Hz) during loudspeaker playback of a representative utterance (\texttt{84-121123-0022}).

\noindent\textbf{Modal frequency response and axis coupling.}
When speech plays through the built-in loudspeaker, acoustic pressure and electromechanical reaction forces excite structural bending modes across the chassis and PCB, coupling speech energy unevenly across gyroscope axes and frequency bands. As shown in Figure~\ref{fig:multirate_physics}(a), vibration couples predominantly into the $x$-axis (18.9\,dB mean SNR, vs.\ 14.1\,dB on $y$ and 3.3\,dB on $z$). Within the dominant axis, the chassis--gyroscope frequency response amplifies voiced harmonics well above the readout noise floor (exceeding 20\,dB spectral SNR in Figures~\ref{fig:multirate_physics}(b,e)), while shaping their relative envelope amplitudes according to the device's modal resonances.

\noindent\textbf{Sampling-rate tiers and aliasing separation.}
Once structural coupling preserves these speech harmonics, the exported sampling rate governs whether their time--frequency trajectories remain separable within the 20--25\,ms short-time windows required to track phoneme transitions~\cite{davis1980comparison,rabiner2007introduction}. Under unthrottled Camera2 metadata delivery at $\approx 1000$\,Hz (Figure~\ref{fig:multirate_physics}(b), $f_{\mathrm{Nyq}} \approx 499$\,Hz), each 25\,ms window contains 25 temporal samples: the fundamental pitch $f_0$ ($\approx 191$\,Hz) and second harmonic $2f_0$ ($\approx 378$\,Hz) are captured unaliased, while $3f_0$ folds once ($f_s - 3f_0 \approx 427$\,Hz) into an open band and $>1$\,kHz harmonics attenuate toward the noise floor under vocal-source roll-off, post-resonance inertia, and gyroscope low-pass filtering~\cite{rabiner2007introduction,bosch_bmi270,st_lsm6dsox}.

By contrast, public motion sensors under platform rate controls suffer severe aliasing self-interference. At the 500\,Hz ceiling of standalone \texttt{SensorManager} on this phone with \texttt{HIGH\_SAMPLING\_RATE\_SENSORS} (Figure~\ref{fig:multirate_physics}(c), $f_{\mathrm{Nyq}} \approx 249$\,Hz), $2f_0$ folds down to $\approx 121$\,Hz and collides with the fundamental pitch band (on 400\,Hz phones, $2f_0$ folds to $\approx 22$\,Hz). Under Android's default 200\,Hz cap or a disabled microphone privacy toggle (Figure~\ref{fig:multirate_physics}(d), $f_{\mathrm{Nyq}} = 100$\,Hz), the 0--1\,kHz speech band collapses into ten overlapping 100\,Hz sub-bands ($10\times$ aliasing pileup).

As shown by the voiced spectral slice at $t = 0.92$\,s (Figure~\ref{fig:multirate_physics}(e)) and the summary metrics on this trace (Figure~\ref{fig:multirate_physics}(f)), folding collisions at lower rates both erase individual harmonic peaks and elevate the inter-harmonic background floor: moving from the 200\,Hz cap to 1000\,Hz Camera2 metadata increases harmonic-to-background contrast (effective harmonic SNR) by 12.4\,dB (from 8.5 to 20.9\,dB, an $18\times$ larger contrast ratio) and nearly quadruples the empirical Shannon spectral capacity $\int_{0}^{f_{\mathrm{Nyq}}}\log_2(1+\mathrm{SNR}(f))\,df$ across the Nyquist band (from 354 to 1370\,bits/s). While coarse whole-word energy envelopes at 200\,Hz suffice for closed-set digit or hotword classification~\cite{michalevsky2014gyrophone,ba2020learning,gao2022inertiear}, continuous speech recognition requires both preserving harmonic trajectories at 800--1000\,Hz and modeling device-specific modal responses and folding rates across handsets (Section~\ref{sec:systemdesign}).

\newcolumntype{L}[1]{>{\raggedright\arraybackslash}p{#1}}
\newcolumntype{R}{>{\raggedright\arraybackslash}X}

\begin{table*}[t!]
\centering
\caption{Threat scenarios during an active Camera2 session, system protection states, and achievable sampling rates.}
\label{tab:threat_scenarios}
\footnotesize
\setlength{\tabcolsep}{4pt}
\renewcommand{\arraystretch}{1.05}
\begin{tabularx}{\textwidth}{@{}L{0.20\textwidth} L{0.165\textwidth} L{0.125\textwidth} L{0.175\textwidth} R@{}}
\toprule
\textbf{Threat Scenario} & \textbf{Host Process Perms.}\textsuperscript{$\dagger$} & \textbf{Privacy Toggles} & \textbf{Standalone \texttt{SensorManager}} & \textbf{Active Camera2 Rate \& Path} \\
\midrule
\textbf{I. Unprivileged-Host Session}
& \texttt{CAMERA}: \ding{51}\newline
  \texttt{HIGH\_SAMPLING}: \ding{55}
& Mic: \textbf{ON}\newline
  Cam: \textbf{ON}
& \textbf{$\le 200$~Hz} \scriptsize(no permission;\newline
  \texttt{FASTEST} throws exception)
& \textbf{800--1000~Hz}\newline
  \scriptsize In-process \texttt{CaptureResult} (Finding~1) \\
\addlinespace[2pt]
\textbf{II. Privacy-Toggled Session}
& \texttt{CAMERA}: \ding{51}\newline
  \texttt{HIGH\_SAMPLING}: \ding{51}
& Mic: \textbf{OFF}\newline
  Cam: \textbf{OFF} / \textbf{ON}
& \textbf{$\le 200$~Hz}\newline
  \scriptsize(Throttled even w/ \ding{51})
& \textbf{800--1000~Hz}\newline
  \scriptsize In-process \texttt{CaptureResult} (Finding~4) \\
\bottomrule
\end{tabularx}
\vspace{1pt}
\begin{minipage}{\textwidth}
\scriptsize \textsuperscript{$\dagger$}Microphone capture is unavailable in both scenarios: \texttt{RECORD\_AUDIO} is ungranted in Scenario~I, and granted in Scenario~II but silenced by the system-wide Mic toggle.
\end{minipage}
\end{table*}
\noindent\textbf{Finding 3: Loudspeaker speech couples into 800--1000\,Hz Camera2 metadata through axis-dependent structural resonances and retains cleanly separable unaliased and single-folded harmonics, whereas 200--500\,Hz sensor streams suffer severe multi-band aliasing collisions.}
\subsection{Behavior Across Sensor-Control Boundaries}
\label{sec:policy_boundaries}

Android applies permission and rate controls to public motion sensors and device-wide privacy controls to camera and microphone access~\cite{android_sensor_rate_limiting,android_privacy_controls}. We test whether these controls constrain Camera2 gyro metadata.

\noindent\textbf{Permission and privacy controls.}
Four of the six evaluated phones implement Android's system-wide camera and microphone privacy toggles, whereas the two Xiaomi/Redmi HyperOS models omit both AOSP toggles entirely (leaving Camera2 and \texttt{SensorManager} permanently unthrottled by system UI controls).

On the four phones implementing both controls, disabling camera access without revoking \texttt{CAMERA} returns blank image buffers in both active and newly opened Camera2 sessions, while the metadata stream retains its 800--1000~Hz sampling rate. Disabling microphone access limits standalone \texttt{SensorManager} delivery to approximately 200~Hz even when the application declares \texttt{HIGH\_SAMPLING\_RATE\_SENSORS}, but does not reduce the Camera2 metadata sampling rate. Under either privacy-control configuration, the metadata continues to capture the speech-induced harmonic structure shown in Section~\ref{sec:acoustic_modulation}.

\noindent\textbf{Finding 4: On the four phones implementing both device-wide privacy controls, the Camera2 gyro metadata stream remains an 800--1000~Hz acoustic observation path after either camera access or microphone access is disabled.}

\noindent\textbf{Cross-application rate propagation.}
Extending intra-\texttt{SensorManager} rate spillover~\cite{shepherd2023side,long2025armour} across API boundaries, we test whether Camera2's high-rate state propagates through the shared motion pipeline (Finding~2) to a concurrent non-camera app. First, a concurrent \texttt{SensorManager} observer declaring \texttt{HIGH\_SAMPLING\_RATE\_SENSORS} receives 400--500~Hz standalone, but jumps to 800--1000~Hz on all six phones whenever a foreground app opens Camera2.

Second, a concurrent observer capped at $\approx$200~Hz (either because it omits \texttt{HIGH\_SAMPLING\_RATE\_SENSORS} or because microphone access is disabled) rises to 400--500~Hz when a foreground Camera2 session opens, occurring on five of six phones without the permission and three of the four toggle-equipped phones (on the fourth, vivo X200 Ultra, it receives 182~Hz as a one-in-five subset of Camera2 timestamps). Across all configurations, opening Camera2 switches the shared sensor pipeline into an 800--1000~Hz operating state that propagates across API and application boundaries due to the lack of per-client rate isolation.

\noindent\textbf{Finding 5: Opening a Camera2 session globally elevates the shared hardware gyroscope rate, boosting concurrent \texttt{SensorManager} listeners from 200~Hz to 400--500~Hz without \texttt{HIGH\_SAMPLING\_RATE\_SENSORS} and from 400--500~Hz to 800--1000~Hz with it.}

Together, these measurements establish that Camera2 metadata forms an application-visible acoustic side channel beyond Android's sensor-control boundaries.

\section{Threat Model}
\label{sec:threat_model}

\noindent\textbf{Adversary Goal.}
The adversary's goal is to obtain an open-vocabulary transcript of continuous speech played through the victim smartphone's built-in loudspeaker, such as remote participants' speech during a speakerphone meeting, online lecture, telehealth consultation, or voice-message playback. When speech plays through the built-in loudspeaker, the resulting mechanical vibrations propagate through the shared handset chassis and modulate the onboard gyroscope~\cite{ba2020learning,anand2021spearphone,sun2023stealthyimu}. The adversary recovers the spoken content exclusively from the timestamped three-axis motion samples exposed during an active Camera2 session.

\noindent\textbf{Adversarial Capabilities and Threat Scenarios.}
The adversary executes in ordinary application user space where \texttt{RECORD\_AUDIO} is not granted or microphone access is disabled. Because speech transcription relies solely on numeric motion metadata, the attack succeeds even when the lens faces an uninformative scene, is physically covered, or returns blank image buffers under Android's camera privacy toggle.
\begin{figure*}[t!]
\centering
\safeincludegraphics[width=0.83 \textwidth]{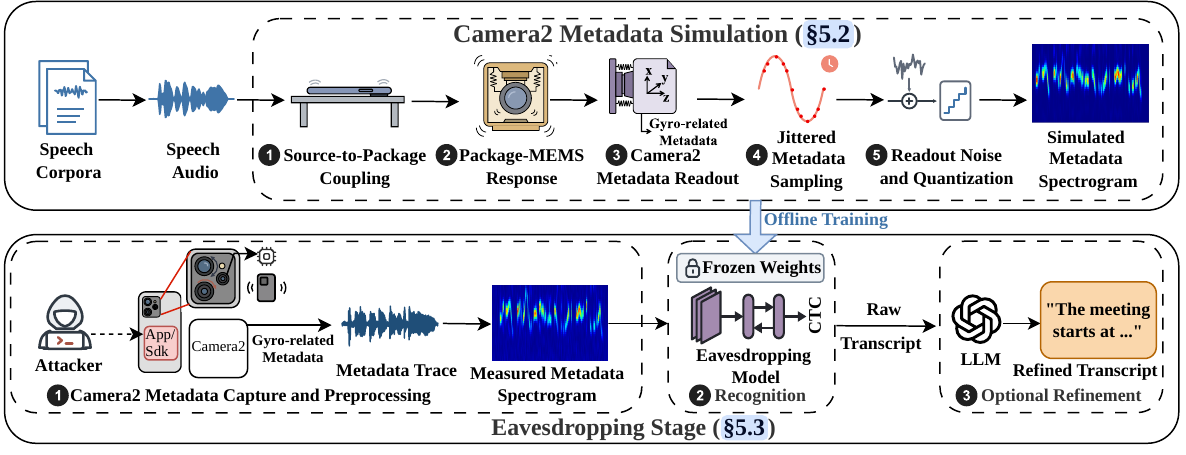}
\caption{\sysname's S2L model-construction and attack flows.}
\label{framework}
\end{figure*}
Collection occurs while a foreground app holds an active rear-camera session (where handset OIS/EIS resides and Android's camera indicator is expected) as remote speech plays over the loudspeaker, such as sharing documents, notes, or physical objects during a meeting, online lecture, or tele-consultation~\cite{jones2015mechanics,licoppe2009collaborative}, or using a camera app during a speakerphone call. Table~\ref{tab:threat_scenarios} defines two progressive in-process scenarios under Android's active protections:
\begin{itemize}[leftmargin=*, nosep]
\item \textbf{Scenario~I (Unprivileged-Host Session).} Because Android grants \texttt{CAMERA} and \texttt{RECORD\_AUDIO} separately, a user may grant \texttt{CAMERA} while denying \texttt{RECORD\_AUDIO} (e.g., sharing desk work or lab setups during a lecture or meeting), or open a camera-only app during a speakerphone call. Neither an untrusted app nor a third-party SDK (e.g., filter, virtual-background, or RTC library) in a benign host~\cite{long2023side} can capture audio, and since benign camera hosts rarely declare \texttt{HIGH\_SAMPLING\_RATE\_SENSORS}~\cite{long2025armour}, \texttt{SensorManager} is capped at 200~Hz (\texttt{SENSOR\_DELAY\_FASTEST} throws a \texttt{SecurityException}~\cite{android_sensor_rate_limiting}); yet active Camera2 callbacks expose 800--1000~Hz gyroscope metadata (Finding~1).
\item \textbf{Scenario~II (Privacy-Toggled Session).} Conversely, even when a two-way meeting or tele-consultation app holds both \texttt{RECORD\_AUDIO} and \texttt{HIGH\_SAMPLING\_RATE\_SENSORS}, the user may disable Android's system-wide microphone toggle to mute local audio---and optionally the camera toggle after showing an item---while listening over the loudspeaker. Turning off the microphone toggle silences audio and forcibly throttles \texttt{SensorManager} back to $\le$200~Hz despite \texttt{HIGH\_SAMPLING\_RATE\_SENSORS}~\cite{android_sensor_rate_limiting,android_privacy_controls} (and turning off the camera toggle blanks video frames); yet the open Camera2 session still streams 800--1000~Hz gyroscope metadata (Finding~4).
\end{itemize}
Via shared-MEMS rate spillover (Findings~2 and~5), a concurrent non-camera app likewise receives this 800--1000~Hz sequence with \texttt{HIGH\_SAMPLING\_RATE\_SENSORS} or its 400--500~Hz subset without it (reaching 32.4\% WAcc in Section~\ref{sec:countermeasures}); our primary evaluation focuses on the unmediated 800--1000~Hz Camera2 stream.

\noindent\textbf{Deployment Workflow.}
Unlike prior continuous-speech motion-sensor attacks that require hours of target-device trace collection~\cite{hu2022accear,liang2022escalated,zhang2023spy}, we operate in a strict \emph{zero-profiling} setting where the adversary has no prior access to the target handset to collect paired traces or estimate its transfer response. Offline, the adversary trains one cross-device recognizer on public speech and simulated Camera2 leakage from our physics-constrained channel family (Section~\ref{sec:systemdesign}). Online, the deployed parser (1)~locates the batched three-axis gyroscope payload among visible \texttt{CaptureResult} vendor keys, (2)~restores chronological sample order and effective sampling rate from its hardware timestamps, and (3)~selects the axis with the highest retained-band energy for transcription.

\section{Source-to-Leakage Learning and \sysname}
\label{sec:systemdesign}

This section formalizes the Source-to-Leakage Learning (S2L) paradigm and instantiates it in \sysname for cross-device speech recognition from Camera2 metadata.

\subsection{S2L Formulation}
\label{sec:s2l-formulation}

Let $p_S$ denote a distribution of labeled source data obtained independently of physical leakage measurements. Each draw $(s,y)\sim p_S$ contains a source signal $s\in\mathcal S$ and recognition target $y\in\mathcal Y$; in \sysname, these are a speech waveform and its transcript. When $s$ excites a physical device instance $d$, the hardware produces leakage $\ell_d = \mathcal H_d(s) + \eta \in \mathcal L$ (with measurement noise $\eta$).

In a zero-profiling cross-device setting where the target hardware $d$ is unseen prior to deployment, the forward operator $\mathcal H_d$ is unknown, creating two fundamental challenges. First, unlike wideband side channels where synthetic denoising can restore audio waveforms for off-the-shelf ASR, inverting a sub-Nyquist inertial trace into clean audio ($\hat s \approx \mathcal H_d^{-1}(\ell_d)$) is severely ill-posed without target-device calibration: sub-Nyquist sampling folds multiple upper spectral bands onto the same baseband frequencies, while $\mathcal H_d$ applies unknown device-specific resonant gains and phase shifts~\cite{hu2022accear,liang2022escalated}. Second, training a recognizer directly on measured pairs $(\ell_d,y)$ from a fixed set of training phones binds the labels $y$ to those phones' specific operators $\mathcal H_d$: recording more speech on the same devices expands $p_S$, yet leaves hardware diversity restricted to the measured handsets.

Our key insight is that while a measured trace $\ell_d$ irreversibly entangles $s$ and $\mathcal H_d$ after collection, the underlying physical process $\ell_d = \mathcal H_d(s) + \eta$ takes $s$ and $\mathcal H_d$ as independent inputs prior to measurement. Unlike spectrogram augmentations~\cite{ko2015audio,park2019specaugment}, which mask or warp features while preserving harmonic bin order, and domain randomization~\cite{tobin2017domain}, which applies unconstrained perturbations, S2L models pre-sampling modal resonances and rate-dependent aliasing folding in $\{\widetilde{\mathcal H}_{\theta}: \mathcal S \to \mathcal L \mid \theta\in\Theta\}$ under $p_\Theta$. During training, S2L independently draws source--target pairs $(s,y) \sim p_S$ and channel parameters $\theta \sim p_\Theta$ to synthesize supervision $(\widetilde{\mathcal H}_{\theta}(s),y)$, optimizing
\begin{equation}
    \phi^\star =
    \arg\min_{\phi}
    \mathbb{E}_{(s,y)\sim p_S}
    \mathbb{E}_{\theta\sim p_{\Theta}}
    \left[
    \ell_{\mathrm{CTC}}\!\left(
    f_{\phi}\!\left(\widetilde{\mathcal H}_{\theta}(s)\right),
    y
    \right)
    \right].
    \label{eq:s2l-objective}
\end{equation}
Because sub-Nyquist folding and modal resonances produce deterministic, structured spectral patterns, optimizing independently over $p_S$ and $p_\Theta$ enables $f_{\phi^\star}$ to map aliased spectrograms directly to transcripts across unseen hardware without inverting $\mathcal H_d$.
\subsection{\sysname Instantiation and Training Distribution}
\label{sec:model-generation}

\sysname instantiates $p_S$ using a public speech corpus and $p_\Theta$ over the Camera2 gyroscope channel family, synthesizing pairs $\bigl(\widetilde{\mathcal H}_{\theta}(s), y\bigr)$ on the fly to train a CNN--BiLSTM--CTC recognizer.

As shown in Figure~\ref{framework}, the forward operator $\widetilde{\mathcal H}_{\theta}$ models five acquisition stages in order: \ding{182}~loudspeaker playback excites the gyroscope package through the chassis; \ding{183}~package and MEMS dynamics shape the vibration response; \ding{184}~Camera2 metadata readout projects motion onto an exported axis; \ding{185}~jittered sub-Nyquist sampling folds speech harmonics; and \ding{186}~readout noise and quantization produce the visible metadata. Parameter supports are derived from camera-vibration studies and commercial OIS/IMU specifications~\cite{long2023side,la2015optical,bosch_bmi270,st_lsm6dsox}; Appendix~\ref{append:model} reports the numerical parameter ranges and implementation details, which remain fixed across all evaluated phones.

\subsubsection{Camera2 Gyro Metadata Simulation}
\label{sec:camera-metadata-simulation}

For the dominant exported gyroscope axis (Section~\ref{sec:metadata-representation}), we express the forward realization $\widetilde{\mathcal H}_{\theta}$ mapping a 16~kHz speech waveform $u$ to a synthetic metadata stream $\widetilde\ell_{\theta}$ as a composite operator:
\begin{equation}
    \widetilde\ell_{\theta}
    =
    \widetilde{\mathcal H}_{\theta}(u)
    =
    \mathcal Q_{\delta_{\theta}}\!\Bigl(
        \mathcal S_{\tau_{\theta}}\!\bigl(\mathcal L_{\theta}(u)\bigr)
        + \eta_{\theta}
    \Bigr).
    \label{eq:s2l-composite}
\end{equation}
This composition traces the five physical acquisition stages in Figure~\ref{framework}.

\noindent\textbf{\ding{182} Acoustic-to-package excitation.}
Acoustic pressure and electromechanical reaction forces from the built-in loudspeaker excite structural bending waves across the handset chassis and PCB. These waves propagate mechanically to the gyroscope package, driving the sensor with the speech spectrum $U(\omega) = \mathcal F\{u\}(\omega)$ on the $F_w = 16$~kHz waveform grid.

\noindent\textbf{\ding{183} Package--MEMS modal dynamics.}
Package vibration couples into the gyroscope sense path through local angular motion, mount asymmetry, cross-axis sensitivity, and structural resonances~\cite{michalevsky2014gyrophone,son2015rocking,tatar2012quadrature}. Following modal analysis~\cite{bk_modal_testing}, each transfer branch $R^{q}_{\theta}(\omega)$ ($q\in\{\mathrm p,\mathrm r\}$) aggregates $M$ underdamped structural modes in complex-conjugate pole-residue form:
\begin{equation}
    R^{q}_{\theta}(\omega)
    =
    \sum_{m=1}^{M}
    \frac{a^{q}_{m,\theta}\,|p_{m,\theta}|^{2}}
         {(j\omega - p_{m,\theta})(j\omega - p_{m,\theta}^{*})},
    \label{eq:gr-poles}
\end{equation}
with poles $p_{m,\theta} = -\sigma_{m,\theta} + j\omega_{d,m,\theta}$, where $\sigma_{m,\theta}>0$ controls modal decay, $\omega_{d,m,\theta}>0$ sets the damped modal frequency, and the signed residues $a^{q}_{m,\theta}$ absorb source-to-package coupling and single-axis projection.

\noindent\textbf{\ding{184} Camera2 readout transfer.}
The electro-mechanical operator $\mathcal L_{\theta}$ combines primary angular-rate transduction ($\mathrm p$, $j\omega/\omega_0$ with $\omega_0 = 2\pi f_{\mathrm{ref}}$) with residual cross-axis displacement, velocity, or acceleration coupling ($\mathrm r$, $\nu_{\theta}\in\{0,1,2\}$), followed by the gyroscope low-pass roll-off $B_{\theta}(\omega)=\bigl[1+(|\omega|/\omega_{c,\theta})^{2p_{\theta}}\bigr]^{-1/2}$:
\begin{equation}
\begin{aligned}
    \mathcal L_{\theta}(u)[n]
    &= \mathcal F^{-1}\!\bigl\{H_{\theta}(\omega)\,U(\omega)\bigr\}[n], \\
    H_{\theta}(\omega)
    &= B_{\theta}(\omega)\!\left[
        \lambda_{\theta}\frac{j\omega}{\omega_0}R^{\mathrm p}_{\theta}(\omega)
        \!+\! (1\!-\!\lambda_{\theta})\!\left(\frac{j\omega}{\omega_0}\right)^{\!\nu_{\theta}}\!R^{\mathrm r}_{\theta}(\omega)
    \right]\!.
\end{aligned}
\label{eq:gr-operator}
\end{equation}
Here, $\lambda_{\theta}$ weights the primary and cross-axis paths, and the filtered spectrum $H_{\theta}(\omega)U(\omega)$ is normalized by its peak magnitude before inverse Fourier transformation.

\noindent\textbf{\ding{185} Irregular metadata sampling.}
Because the Camera2 HAL batches sensor-hub readings into $\approx 30$~fps \texttt{CaptureResult} payloads, sorting by hardware timestamp produces a sub-Nyquist sequence at nominal rate $F_{\theta} \ll F_w$. Sampling below the speech bandwidth folds harmonics above $F_{\theta}/2$ into the baseband, shifting folded contours across handsets with different metadata rates (e.g., 799~Hz vs.\ 998~Hz). To capture both cross-model rate variation and FIFO timing jitter $\delta\tau_i \sim \mathcal N(0,\sigma_{\tau,\theta}^{2})$ around nominal period $1/F_{\theta}$, $\mathcal S_{\tau_{\theta}}$ evaluates $z = \mathcal L_{\theta}(u)$ along arrival times $\{t_i\}$ aligned to the $1/F_w$ grid:
\begin{equation}
\begin{aligned}
    \mathcal S_{\tau_{\theta}}(z)[i] &= z[t_i F_w], \\
    t_i &= t_{i-1} + \max\!\left(\frac{1}{F_w},\,\operatorname{round}_{1/F_w}\!\left(\frac{1}{F_{\theta}} + \delta\tau_i\right)\right),
\end{aligned}
\label{eq:ms-readout}
\end{equation}
with $t_0 = 0$ and $\operatorname{round}_{1/F_w}(v) = F_w^{-1}\lfloor F_w v \rceil$.

\noindent\textbf{\ding{186} Readout noise and quantization.}
Finally, adding zero-mean sensor readout noise $\eta_{\theta}[i] \sim \mathcal N(0,\sigma_{\eta,\theta}^{2})$ and applying finite-resolution scalar quantization $\mathcal Q_{\delta_{\theta}}$ with step size $\delta_{\theta}$ in Equation~\eqref{eq:s2l-composite} produces the visible Camera2 metadata stream $\widetilde\ell_{\theta}$.

\subsubsection{Shared Metadata Representation}
\label{sec:metadata-representation}

Simulated and measured traces enter the recognizer through an identical time--frequency representation (Appendix~\ref{append:representation}). For measured three-axis metadata, hardware timestamps restore chronological sample order (and determine the effective rate $\widehat F$ for diagnostics). After removing the per-trace DC offset, each axis $q\in\{x,y,z\}$ is converted directly into a short-time Fourier transform (STFT) magnitude spectrogram $S_q(f,t)$ without time-domain resampling. Because human hand motion and gravity drift concentrate at low frequencies whereas loudspeaker coupling excites higher-frequency structural modes, we discard the lowest frequency bins to form the retained bin set $\mathcal B$, select the axis $q^\star$ maximizing mean-square magnitude over $\mathcal B$, and apply per-trace min--max normalization:
\begin{equation}
\begin{aligned}
    q^\star &= \arg\max_{q\in\{x,y,z\}} \frac{1}{T|\mathcal B|}\sum_{t=1}^{T}\sum_{f\in\mathcal B}S_q(f,t)^2,\\
    \widetilde S(f,t) &= \frac{S_{q^\star}(f,t)-\min S_{q^\star}}{\max S_{q^\star}-\min S_{q^\star}+\epsilon}.
\end{aligned}
\label{eq:representation-norm}
\end{equation}

\subsubsection{Learning Over the Camera2 Channel Family}
\label{sec:model-training}

\sysname trains a CNN--BiLSTM--CTC recognizer to map normalized metadata spectrograms to lowercase, punctuation-free character sequences~\cite{graves2006connectionist}. Convolutional blocks extract local harmonic contours from the aliased spectrogram, while bidirectional LSTM layers aggregate sequential context across frames. Connectionist Temporal Classification (CTC) sums over frame-level alignments that collapse to the target character sequence, enabling end-to-end training on utterance-level pairs without frame-level alignment labels. During training, modal response realizations are composed with metadata-sampling processes redrawn at every epoch, and greedy CTC decoding produces the transcript at inference time. The feature representation, network architecture, training schedule, and model checkpoint are shared across all evaluated phones (Appendix~\ref{append:model}).

\subsection{Profiling-Free Deployment}
\label{sec:eavesdropping-framework}

During deployment on an unseen phone (bottom flow of Figure~\ref{framework}), \sysname runs three stages: \textbf{(1)~Camera2 metadata capture and preprocessing} identifies the gyro vendor field by key name and three-axis payload structure (Section~\ref{sec:measurement_method}), orders samples by hardware timestamp, and selects the dominant axis $q^\star$ for the normalized spectrogram $\widetilde S(f,t)$ (Equation~\eqref{eq:representation-norm}); \textbf{(2)~Eavesdropping model recognition} decodes $\widetilde S(f,t)$ into a raw character sequence with the pre-trained greedy CTC recognizer; and \textbf{(3)~Optional refinement} feeds this character stream to a fixed text-only LLM prompt (Appendix~\ref{append:llm}) to resolve phonetic substitutions into a sentence transcript.

\section{Evaluation Results}
\label{sec:evaluation}

We evaluate \sysname across the six phones exposing 800--1000~Hz Camera2 metadata (Section~\ref{sec:measurement}) to answer three research questions:
\begin{itemize}[leftmargin=*, nosep]
\item \textbf{RQ1: Zero-profiling cross-device portability.} How accurately does a single simulation-trained \sysname recognizer transcribe continuous speech across unseen handsets compared with 24-hour per-device fine-tuned ISpyU models?
\item \textbf{RQ2: Physical channel deconstruction.} How do gyroscope modal-response (\textbf{GR}) and sub-Nyquist metadata-sampling (\textbf{MS}) modeling each contribute to simulated spectrogram fidelity and cross-device recognition?
\item \textbf{RQ3: Operating sensitivity and corpus scaling.} How sensitive is zero-profiling speech recovery to physical operating conditions, linguistic context, and offline source-corpus scale?
\end{itemize}

\subsection{Evaluation Protocol}
\label{sec:eval_setup}

\noindent\textbf{Devices and dataset.}
For each test recording, a foreground app opens the rear-camera preview (Section~\ref{sec:measurement}) and logs the timestamped gyro metadata stream. Unless stated otherwise, each phone rests alone on a rigid tabletop playing speech from its internal loudspeaker via \texttt{STREAM\_MUSIC} at 100\% OS volume~\cite{anand2021spearphone,zhang2023spy}. Table~\ref{tab:device_list} lists the six evaluation phones (X300s, X300P, X200U, 17TP, M8PA, and K90M), all on Android~16. Following standard ASR and continuous-speech side-channel benchmarks~\cite{panayotov2015librispeech,hu2022accear,zhang2023spy}, we synthesize training leakage from LibriSpeech \textit{train-960} ($\sim$960~hours of 16-kHz real-speaker English speech) and evaluate on 816 speaker-disjoint utterances (15{,}293 reference words) from 16 randomly sampled unseen speakers (eight male, eight female) in \textit{dev-clean}, played over each phone's loudspeaker to collect 4{,}896 measured Camera2 test traces. Unless noted otherwise, experiments use the STFT representation, greedy CTC recognizer, and SpecAugment~\cite{park2019specaugment} training setup in Appendix~\ref{append:implement}.

\begin{table}[t!]
\centering
\caption{Evaluated 800--1000~Hz Camera2 handsets.}
\small
\setlength{\tabcolsep}{4pt}
\begin{threeparttable}
\begin{tabular}{l c c}
\toprule
\textbf{Model} & \textbf{Metadata Rate} & \textbf{Release Year} \\
\midrule
vivo X300s (X300s)          & $\approx$ 998~Hz & 2026 \\
vivo X300 Pro (X300P)       & $\approx$ 959~Hz & 2025 \\
vivo X200 Ultra (X200U)     & $\approx$ 910~Hz & 2025 \\
Xiaomi 17T Pro (17TP)       & $\approx$ 801~Hz & 2026 \\
HONOR Magic8 Pro Air (M8PA) & $\approx$ 801~Hz & 2026 \\
Redmi K90 Max (K90M)        & $\approx$ 799~Hz & 2026 \\
\bottomrule
\end{tabular}
\end{threeparttable}
\label{tab:device_list}
\end{table}

\noindent\textbf{Recognition metrics.}
Under unit-cost Levenshtein edit distance over $N$ reference words on a phone, we define corpus-level word error rate $\mathrm{WER} = (S+D+I)/N$, decomposed into substitution ($\mathrm{Sub.}=S/N$), deletion ($\mathrm{Del.}=D/N$), and insertion ($\mathrm{Ins.}=I/N$) rates, and word accuracy as
\begin{equation}
    \mathrm{WAcc} = \max(0,\, 1 - \mathrm{WER}),
    \label{eq:metamic-accuracy}
\end{equation}
with character accuracy ($\mathrm{CAcc}$) defined analogously over reference characters. All rates and accuracies are reported as percentages, weighting all six phones equally in multi-device means.

\subsection{Cross-Device Continuous-Speech Recognition}
\label{sec:metamic_perf}

\providecommand{\CDOod}[1]{\fontsize{8}{8}\selectfont #1}

\noindent\textbf{Baseline.}
Among other sub-kilohertz inertial attacks, closed-set classifiers recognize only digits or hotwords~\cite{michalevsky2014gyrophone,ba2020learning,gao2022inertiear}, while audio-reconstruction systems target human listening (scored via spectral distortion or human listeners, with no transcription stage) and require paired device--audio GAN training~\cite{hu2022accear,liang2022escalated}, leaving ISpyU~\cite{zhang2023spy} as the sole automated transcript baseline. In ISpyU's design~\cite{zhang2023spy}, Stage~1 decimates 16-kHz LibriSpeech by a uniform integer factor ($32\times$ at 500~Hz) under a flat response, relying on Stage-2 target-phone traces to adapt to hardware resonances and rates. With ISpyU unreleased, we reproduce its two-stage procedure under our matched Camera2 setup (Appendix~\ref{append:ispyu-reproduction}): Stage~1 pretrains on 281{,}241 \textit{train-960} utterances via uniform $16\times$ decimation (8.2\% WAcc; adding variable-rate sampling gives \textbf{MS} at 30.6\% in Table~\ref{tab:metamic_ablation}), and Stage~2 fine-tunes on 7{,}681 real traces ($\sim$24~hours) per phone $d_i$.

\noindent\textbf{Recognition results.}
Let $S_{i,j}$ denote the WAcc of the ISpyU model profiled on phone $d_i$ and evaluated on $d_j$, and $S_{\sysname,j}$ the WAcc of \sysname on $d_j$, with diagonal entries ($i=j$) measuring same-device accuracy and cross-device means averaging over unseen handsets (overall ISpyU cross-device mean averaging all six columns):
\begin{equation}
    \mathrm{CD}_{\mathrm{ISpyU}}(i) = \frac{1}{5}\sum_{j\neq i}S_{i,j},
    \quad
    \mathrm{CD}_{\sysname} = \frac{1}{6}\sum_{j=1}^{6}S_{\sysname,j}.
\label{eq:metamic-ood}
\end{equation}
In Table~\ref{tab:ispyu_asr}, using zero target-phone traces, one \sysname checkpoint achieves 52.1\% mean WAcc (45.0--61.8\%) and 74.3\% mean CAcc (69.3--80.7\%); as shown in Table~\ref{tab:wacc_examples} (Appendix~\ref{append:wacc_examples}), residual raw errors around 50\% WAcc are mostly local character substitutions preserving legibility (e.g., \textit{``i ll be glad to sty sir she replied''} for \textit{``ill be glad to try sir he replied''} at 50\% WAcc). Across the six fine-tuned ISpyU models, cross-device mean $\mathrm{CD}_{\mathrm{ISpyU}}(i)$ averages 25.9\% WAcc (21.1--30.0\%, peaking on M8PA), so \sysname improves cross-device WAcc by 101.4\% (+26.2 percentage points, from unrounded scores). Pooling the same 24-hour budget across all five non-target phones in a leave-one-out baseline (Table~\ref{tab:ispyu_lodo}, Appendix~\ref{append:ispyu-reproduction}) reaches only 29.7\% mean WAcc (54.2\% CAcc), trailing \sysname by 22.4 percentage points (+75.5\%). \sysname also surpasses ISpyU's 24-hour same-device diagonal models (36.6\% mean, up to 57.3\% on X300s) on every phone (+15.5 percentage points).

\begingroup
\setlength{\tabcolsep}{2pt}
\setlength{\aboverulesep}{0pt}
\setlength{\belowrulesep}{0pt}
\renewcommand{\arraystretch}{1.15}
\begin{table}[t!]
\centering
\caption{Cross-device speech recognition (WAcc, \%; ISpyU same-device: 36.6\%, cross-device: 25.9\%).}
\label{tab:ispyu_asr}
\footnotesize
\begin{tabular}{@{}l *{6}{>{\centering\arraybackslash}p{0.84cm}} >{\centering\arraybackslash}p{0.84cm}@{}}
\toprule
 & \multicolumn{6}{c}{\textbf{ISpyU~\cite{zhang2023spy} (Profiled Handset)}} & \textbf{Ours} \\
\cmidrule(lr){2-7}\cmidrule(l){8-8}
\textbf{Eval.} & \textbf{X300s} & \textbf{X300P} & \textbf{X200U} & \textbf{17TP} & \textbf{M8PA} & \textbf{K90M} & \textbf{--} \\
\midrule
X300s & \ScoreCell{57.3} & \ScoreCell{49.0} & \ScoreCell{46.6} & \ScoreCell{35.3} & \ScoreCell{42.2} & \ScoreCell{43.1} & \ScoreCell{61.8} \\
X300P & \ScoreCell{31.4} & \ScoreCell{35.8} & \ScoreCell{29.6} & \ScoreCell{19.9} & \ScoreCell{27.2} & \ScoreCell{23.8} & \ScoreCell{45.0} \\
X200U & \ScoreCell{19.2} & \ScoreCell{19.0} & \ScoreCell{32.5} & \ScoreCell{16.9} & \ScoreCell{23.4} & \ScoreCell{21.8} & \ScoreCell{50.6} \\
17TP  & \ScoreCell{18.1} & \ScoreCell{18.7} & \ScoreCell{22.5} & \ScoreCell{29.9} & \ScoreCell{28.6} & \ScoreCell{28.0} & \ScoreCell{51.1} \\
M8PA  & \ScoreCell{18.0} & \ScoreCell{19.1} & \ScoreCell{22.5} & \ScoreCell{19.2} & \ScoreCell{31.2} & \ScoreCell{24.9} & \ScoreCell{48.9} \\
K90M  & \ScoreCell{18.6} & \ScoreCell{17.7} & \ScoreCell{23.4} & \ScoreCell{19.9} & \ScoreCell{28.6} & \ScoreCell{33.0} & \ScoreCell{55.3} \\
\midrule
\textbf{CD Mean} & 21.1 & 24.7 & 28.9 & 22.2 & 30.0 & 28.3 & \textbf{52.1} \\
\bottomrule
\end{tabular}
\end{table}
\endgroup

\begin{figure}[t!]
    \centering
    \safeincludegraphics[width=\columnwidth]{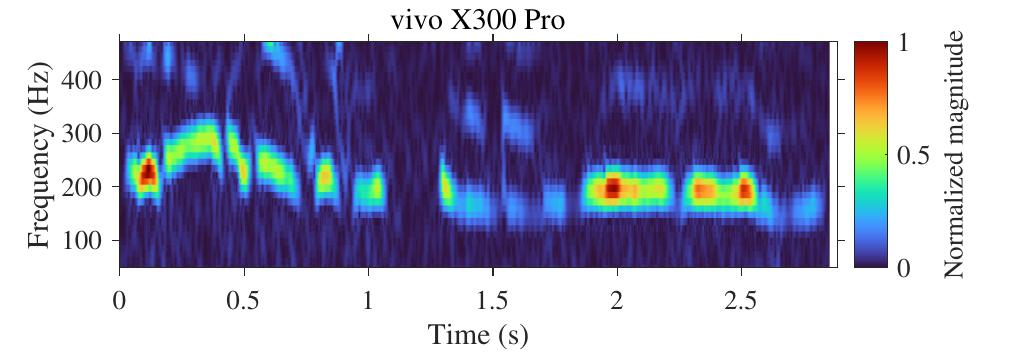}\\[2pt]
    \safeincludegraphics[width=\columnwidth]{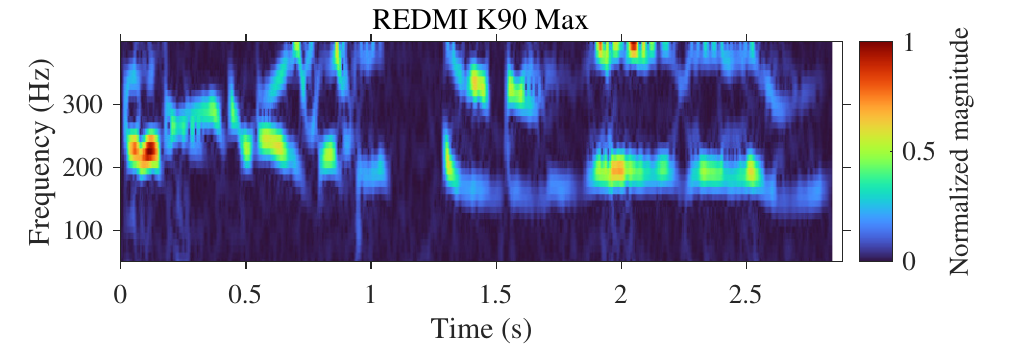}
    \caption{Source-matched Camera2 spectrograms: K90 Max retains clearer harmonics than X300 Pro.}
    \label{fig:device_observability}
\end{figure}

The accuracy matrix in Table~\ref{tab:ispyu_asr} separates per-handset signal fidelity (comparing evaluation rows) from cross-device transferability (comparing model columns). Comparing evaluation rows, every recognizer (both \sysname at 61.8\% and all six ISpyU models at 35.3--57.3\%) achieves its highest accuracy on X300s, whose strong gyroscope chassis coupling and $\approx 499$~Hz Nyquist bandwidth preserve unaliased $f_0$ and $2f_0$ harmonic contours (Section~\ref{sec:acoustic_modulation}). Comparing K90 Max ($\approx 799$~Hz, 55.3\% WAcc) against X300 Pro ($\approx 959$~Hz, 45.0\% WAcc) confirms that chassis coupling co-governs accuracy alongside rate: Figure~\ref{fig:device_observability} shows K90 Max preserves clearer harmonic contours above the noise floor.

Comparing model columns in Table~\ref{tab:ispyu_asr} illustrates how per-phone adaptation specializes to $\mathcal H_{d_i}$ while remaining capped on the diagonal. Even on 998~Hz X300s (matching $16\times$ decimation within 0.2\%), \sysname beats 24-hour same-device ISpyU (61.8\% vs.\ 57.3\%), which drops to 31.4\% on X300P (959~Hz) and 18.0--19.2\% on the other four phones; even across the three rate-matched $\approx 800$~Hz handsets (17TP, M8PA, and K90M), 24-hour adaptation reaches only 29.9--33.0\% same-device and 19.2--28.6\% cross-device WAcc (vs.\ 48.9--55.3\% for \sysname). Because ISpyU's uniform-decimation Stage-1 pretraining omits modal resonance and sampling-rate variation (reaching only 8.2\% WAcc in Table~\ref{tab:metamic_ablation}), even 24~hours of target-phone traces (7{,}681 utterances) cannot bridge this large physical gap, falling short of full-corpus lexical coverage on the profiling phone while overfitting to its $\mathcal H_{d_i}$. Exposing the recognizer to the full 960-hour corpus (281{,}241 utterances) across $p_\Theta$ enables simultaneous generalization across diverse modal responses and sampling rates.

\subsection{Deconstructing S2L: Operator Ablation and Spectrogram Fidelity}
\label{sec:sim_coverage}

To explain why \sysname generalizes without target traces, RQ2 factors $\widetilde{\mathcal H}_\theta$ into gyroscope modal-response (\textbf{GR}, Equations~\eqref{eq:gr-poles}--\eqref{eq:gr-operator}) and sub-Nyquist metadata-sampling (\textbf{MS}, Equation~\eqref{eq:ms-readout}) stages in a $2\times2$ ablation across recognition accuracy (Table~\ref{tab:metamic_ablation}) and spectrogram fidelity (Table~\ref{tab:metamic_feat_sim}).

\begin{table}[t!]
\centering
\caption{End-to-end speech-recognition ablation across simulator variants (raw WAcc, \%).}
\small
\renewcommand{\arraystretch}{1.1}
\setlength{\aboverulesep}{0pt}
\setlength{\belowrulesep}{0pt}
\setlength{\tabcolsep}{0pt}
\begin{tabular}{@{} >{\raggedright\arraybackslash}p{0.21\columnwidth} | >{\centering\arraybackslash}p{0.29\columnwidth} | *{3}{>{\centering\arraybackslash}p{0.163\columnwidth}} @{}}
\toprule
\textbf{Device} & \textbf{Baseline} & \multicolumn{3}{c}{\textbf{Ours (Recognition WAcc)}} \\
\cmidrule(lr){2-2}\cmidrule(l){3-5}
\footnotesize(WAcc, \%) & \textbf{ISpyU-sim~\cite{zhang2023spy}} & \textbf{GR} & \textbf{MS} & \textbf{GR\&MS} \\
\midrule
X300s & 15.5 & 54.0 & 35.6 & \textbf{61.8} \\
X300P &  6.7 & 39.7 & 23.6 & \textbf{45.0} \\
X200U & 10.8 & 30.7 & 25.1 & \textbf{50.6} \\
17TP  &  4.6 & 39.3 & 31.9 & \textbf{51.1} \\
M8PA  &  5.9 & 41.5 & 33.4 & \textbf{48.9} \\
K90M  &  5.8 & 43.6 & 33.8 & \textbf{55.3} \\
\midrule
Mean  &  8.2 & 41.5 & 30.6 & \textbf{52.1} \\
\bottomrule
\end{tabular}
\label{tab:metamic_ablation}
\end{table}

\subsubsection{Ablation Variants}
\label{sec:sim_variants}

The baseline simulator, \textbf{ISpyU-sim}, reproduces ISpyU's synthetic pretraining operator following~\cite{zhang2023spy}, decimating 16-kHz LibriSpeech at a fixed nominal factor ($16\times$ for the 1-kHz Camera2 band, matching its $32\times$ factor for 500~Hz sensors). \textbf{GR} introduces our multi-mode chassis--MEMS transfer operator before fixed $16\times$ decimation; \textbf{MS} adds our variable-rate, jittered metadata-sampling operator directly onto ISpyU-sim; and \textbf{GR\&MS} composes both physical stages into $\widetilde{\mathcal H}_\theta$. All four variants train the identical acoustic-to-character CTC recognizer for 50 epochs under the same SpecAugment schedule (Appendix~\ref{append:sim_variants}), so downstream WAcc directly measures the speech cues preserved by each simulator.

\subsubsection{Downstream Speech-Recognition Ablation}
\label{sec:ablation_task_utility}

Table~\ref{tab:metamic_ablation} shows that both physical stages are essential for cross-device transcription. Even with SpecAugment, fixed-decimation ISpyU-sim reaches only 8.2\% mean WAcc (reaching 15.5\% on 998~Hz X300s, where $16\times$ decimation matches the hardware rate, but collapsing to 4.6--5.9\% on the 800~Hz phones). Introducing \textbf{MS} alone raises mean WAcc nearly four-fold to 30.6\% (and by $5.7\text{--}6.9\times$ to 31.9--33.8\% on the 800~Hz handsets) by exposing convolutional feature extractors to rate-dependent harmonic folding across $F_\theta$. Introducing \textbf{GR} alone raises mean WAcc five-fold to 41.5\% (30.7--54.0\%), showing that spanning multi-mode chassis--MEMS resonances and damping profiles is critical for sequence decoding across every phone. Combining both stages in \textbf{GR\&MS} reaches 52.1\% mean WAcc and outperforms every partial variant on every handset, gaining +15.5--26.2 percentage points over \textbf{MS} and +5.3--19.9 percentage points over \textbf{GR}.

\subsubsection{Spectrogram-Level Physical Fidelity}
\label{sec:fidelity_results}

To verify at the spectrogram level that these gains stem from closer physical modeling, we measure how closely realizations drawn from each simulator family match real Camera2 spectrograms. Across 50 randomly sampled held-out utterances per phone, we evaluate each simulator ($K=32$ prior draws for stochastic variants, evaluating variable-rate MS/GR\&MS at its timestamp-derived rate $\widehat F$; Appendix~\ref{append:fidelity-protocol}) and record the peak 2D Pearson correlation ($r_{\mathrm{spec}}$) with $\widetilde S(f,t)$ from Equation~\eqref{eq:representation-norm}.

Table~\ref{tab:metamic_feat_sim} reveals two rate regimes behind both stages' complementarity and their distinct rankings under pixel-wise $r_{\mathrm{spec}}$ versus pooled CNN decoding in Table~\ref{tab:metamic_ablation}. On the 910--998~Hz phones (X300s, X300P, and X200U), fixed $16\times$ decimation already places folded harmonics near their true frequency bins, so modal resonance shaping (\textbf{GR}) raises $r_{\mathrm{spec}}$ from 0.468--0.543 to 0.614--0.820 whereas flat-spectrum \textbf{MS} stays at 0.540--0.554. On the 800~Hz phones (17TP, M8PA, and K90M), 1-kHz versus 800-Hz folding places upper harmonics in shifted frequency bins; because pixel-wise correlation harshly penalizes bin shifts (unlike CNN pooling in Table~\ref{tab:metamic_ablation}), fixed-rate ISpyU-sim (0.157--0.193; 0.341 six-phone mean) and \textbf{GR} (0.218--0.267; 0.494 six-phone mean) yield low $r_{\mathrm{spec}}$, whereas \textbf{MS} aligns the folded harmonic tracks and boosts $r_{\mathrm{spec}}$ by $3.3\text{--}3.7\times$ to 0.576--0.643 (0.578 six-phone mean). Composing both stages in \textbf{GR\&MS} covers both regimes and leads on every handset, achieving 0.702--0.853 across all six phones (0.805 mean; on 998~Hz X300s, GR already reaches 0.813 vs.\ 0.814 as $16\times$ decimation matches its rate).
\begin{table}[t!]
\centering
\caption{Spectrogram fidelity to measured hardware across simulator variants (2D correlation $r_{\mathrm{spec}}$).}
\small
\renewcommand{\arraystretch}{1.1}
\setlength{\aboverulesep}{0pt}
\setlength{\belowrulesep}{0pt}
\setlength{\tabcolsep}{0pt}
\begin{tabular}{@{} >{\raggedright\arraybackslash}p{0.21\columnwidth} | >{\centering\arraybackslash}p{0.29\columnwidth} | *{3}{>{\centering\arraybackslash}p{0.163\columnwidth}} @{}}
\toprule
\textbf{Device} & \textbf{Baseline} & \multicolumn{3}{c}{\textbf{Ours (Fidelity $r_{\mathrm{spec}}$)}} \\
\cmidrule(lr){2-2}\cmidrule(l){3-5}
\footnotesize($r_{\mathrm{spec}}$) & \textbf{ISpyU-sim~\cite{zhang2023spy}} & \textbf{GR} & \textbf{MS} & \textbf{GR\&MS} \\
\midrule
X300s & 0.543 & 0.813 & 0.540 & \textbf{0.814} \\
X300P & 0.515 & 0.820 & 0.545 & \textbf{0.853} \\
X200U & 0.468 & 0.614 & 0.554 & \textbf{0.702} \\
17TP  & 0.171 & 0.229 & 0.610 & \textbf{0.812} \\
M8PA  & 0.193 & 0.267 & 0.643 & \textbf{0.826} \\
K90M  & 0.157 & 0.218 & 0.576 & \textbf{0.825} \\
\midrule
Mean  & 0.341 & 0.494 & 0.578 & \textbf{0.805} \\
\bottomrule
\end{tabular}
\label{tab:metamic_feat_sim}
\end{table}

\begin{figure*}[t!]
    \centering
    \subfloat[Per-handset WAcc \& tabletop CAcc.]{%
        \safeincludegraphics[width=0.32\textwidth]{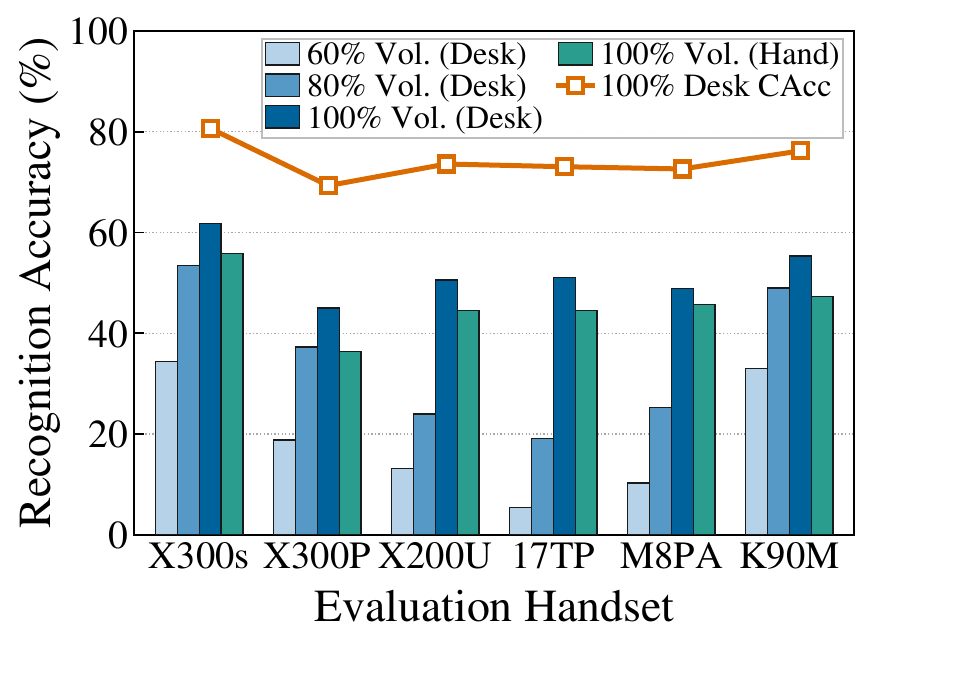}%
        \label{fig:metamic_settings}%
    }\hfill
    \subfloat[WER (bars) \& content recall (line).]{%
        \safeincludegraphics[width=0.32\textwidth]{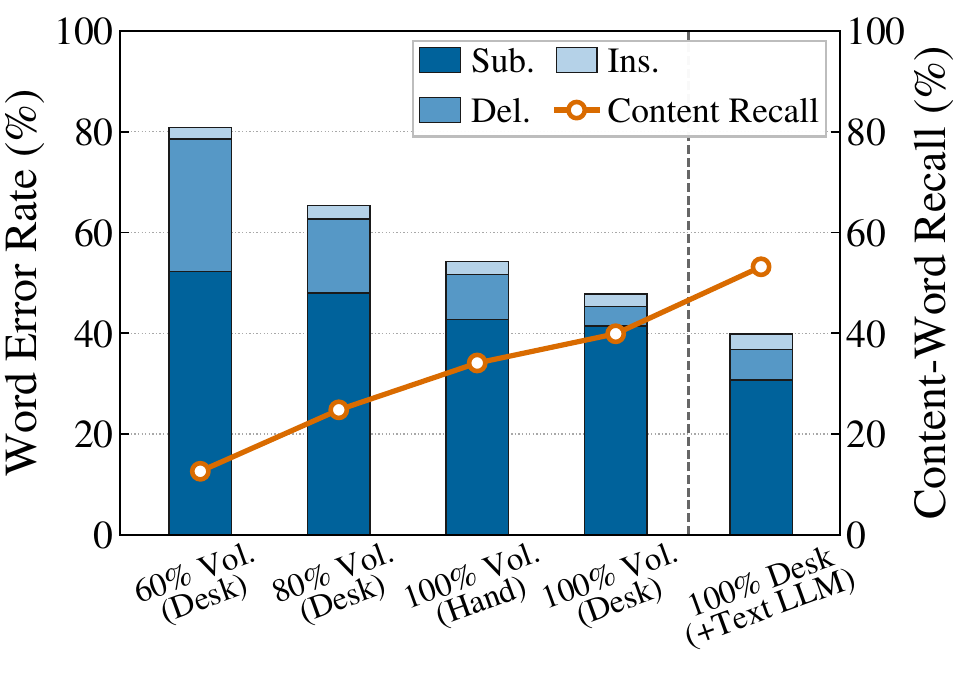}%
        \label{fig:metamic_error_breakdown}%
    }\hfill
    \subfloat[Accuracy by speaker gender.]{%
        \safeincludegraphics[width=0.32\textwidth]{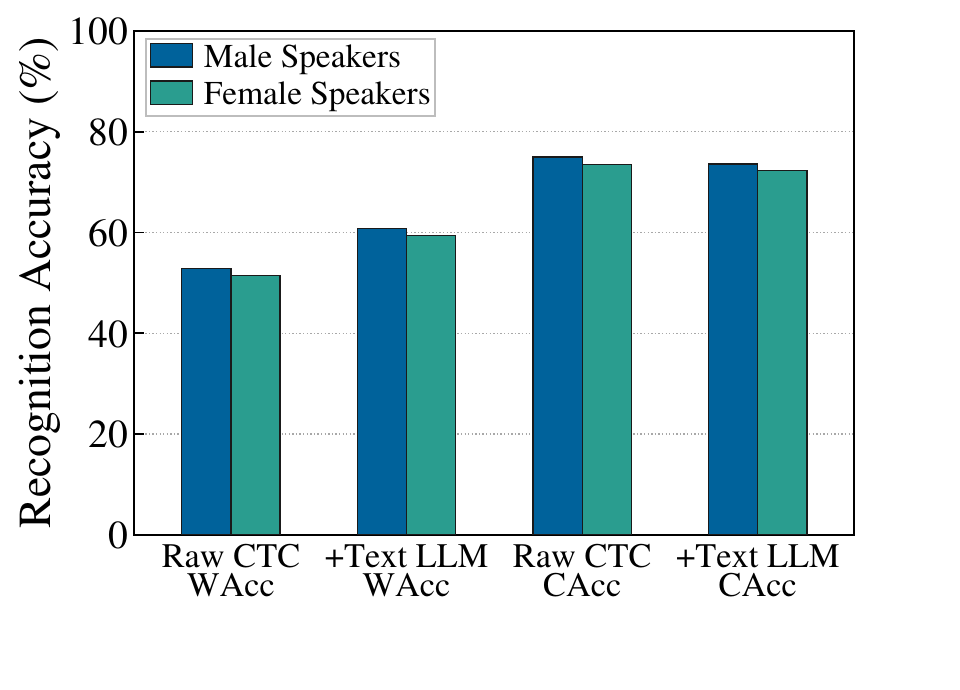}%
        \label{fig:metamic_gender}%
    }
    \caption{\sysname performance across operating conditions, word-error decompositions and content-word recall, and speaker genders (per-handset in (a); six-phone means in (b,c)).}
    \label{fig:metamic_overview}
\end{figure*}
\begin{figure*}[t!]
    \centering
    \subfloat[Accuracy by utterance length.]{%
        \safeincludegraphics[width=0.32\textwidth]{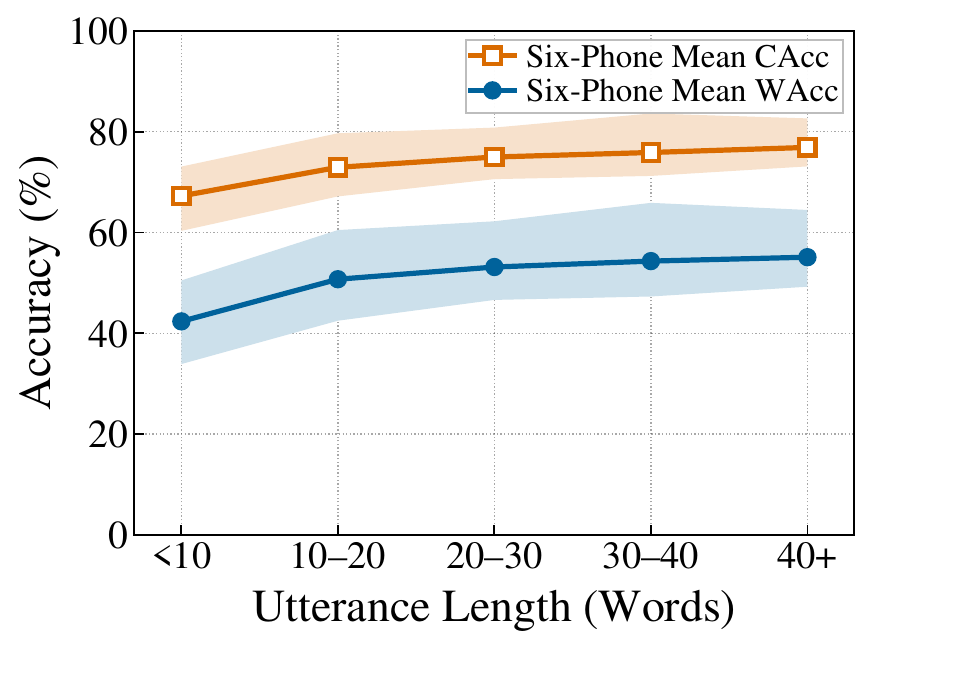}%
        \label{fig:metamic_length}%
    }\hfill
    \subfloat[Utterance-level WAcc CCDFs.]{%
        \safeincludegraphics[width=0.32\textwidth]{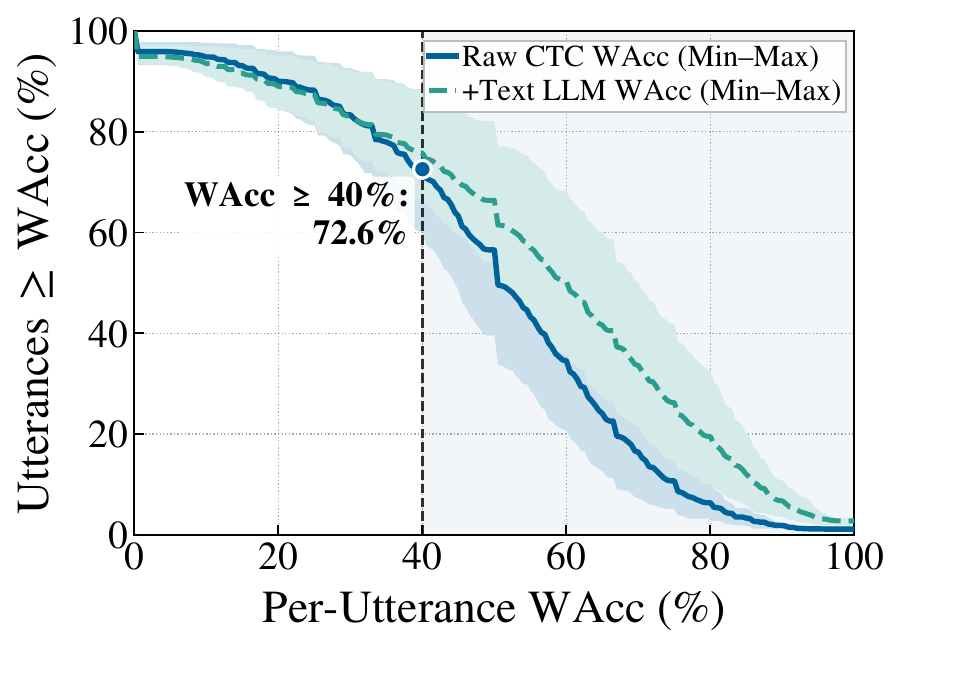}%
        \label{fig:metamic_accuracy_cdf}%
    }\hfill
    \subfloat[Source-corpus scaling in $p_S$.]{%
        \safeincludegraphics[width=0.32\textwidth]{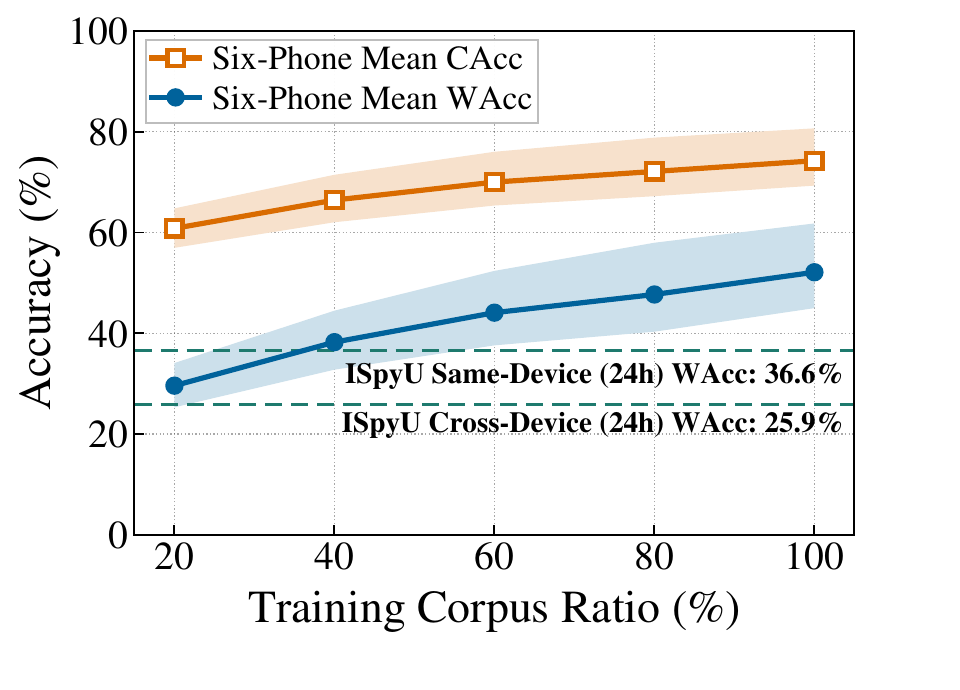}%
        \label{fig:metamic_training_scale}%
    }
    \caption{\sysname recognition accuracy across reference utterance lengths, utterance-level complementary CDFs, and scaling with labeled source-corpus fractions in $p_S$ (six-phone means and min--max bands).}
    \label{fig:metamic_factors_ext}
\end{figure*}
\subsection{Physical Envelope, Contextual Recovery, and Scaling}
\label{sec:metamic_factors}

\subsubsection{Physical Operating Envelope and Error Regimes}
\label{sec:eval_rate_tiers}

We first evaluate two physical operating factors: excitation amplitude (volume and grip) and speaker pitch (across speaker genders). For excitation amplitude (Figures~\ref{fig:metamic_settings} and~\ref{fig:metamic_error_breakdown}, left axis), word errors follow two regimes separated by the gyroscope readout-noise and quantization floor ($\sigma_{\eta,\theta}, \delta_\theta$). At 100\% tabletop volume (52.1\% WAcc, 74.3\% CAcc), chassis vibrations stay well above this floor: whole-word deletions ($\mathrm{Del.}=3.9\%$) and insertions ($\mathrm{Ins.}=2.5\%$) are minimal, and 86.6\% of errors are phonetic substitutions ($\mathrm{Sub.}=41.4\%$) from single-fold aliasing. Holding the phone in hand at 100\% volume remains within this above-threshold regime, maintaining 45.7\% mean WAcc (36.4--55.8\%) and 68.0\% CAcc with only 9.0\% deletions ($\mathrm{Sub.}=42.7\%$), as palm damping moderately attenuates resonant peaks while discarding the lowest STFT bins filters out low-frequency grip motion. By contrast, lowering tabletop volume to 80\% and 60\% pushes weak syllables into the ($\sigma_{\eta,\theta}, \delta_\theta$) floor, inflating deletions nearly seven-fold from 3.9\% to 14.7\% and 26.3\% ($\mathrm{Sub.}=48.0\%$ and $52.3\%$) and reducing mean WAcc to 34.7\% and 19.2\% (CAcc: 58.6\% and 42.7\%, with strongly coupled X300s and K90M retaining 34.4\% and 33.0\% WAcc at 60\% volume versus 5.5--18.8\% on the other four models).

Figure~\ref{fig:metamic_gender} further tests whether speaker gender induces pitch bias. Although adult male and female voices span distinct fundamental-frequency ranges ($\approx$85--180~Hz vs.\ 165--255~Hz~\cite{rabiner2007introduction}), the 400--500~Hz Camera2 Nyquist band retains separable low-order harmonics across both ranges while $p_S$ and $p_\Theta$ (100--1000~Hz) cover both pitch bands. Consequently, male and female means stay within 1.4 points across all metrics (52.8\% (46.2--58.5\%) vs.\ 51.4\% (42.7--65.0\%) raw WAcc, 75.0\% vs.\ 73.6\% CAcc, and 60.8\% vs.\ 59.4\% WAcc with LLM refinement; Section~\ref{sec:error_decoupling}), confirming consistent performance across speaker genders.
\subsubsection{Contextual Disambiguation and Semantic Exposure}
\label{sec:error_decoupling}

Because high-SNR errors are local phonetic substitutions rather than missing words ($\mathrm{Del.}=3.9\%$, $\mathrm{CAcc}=74.3\%$), sequential context at both the sequence-decoding and sentence levels substantially amplifies semantic exposure. At the sequence-decoding level (Figure~\ref{fig:metamic_length}), longer utterances stabilize per-trace spectrogram normalization (Equation~\eqref{eq:representation-norm}) and supply richer cross-word phonotactic context for the BiLSTM layers to disambiguate aliased phonemes: under greedy CTC decoding, six-phone mean WAcc rises monotonically from 42.4\% (67.3\% CAcc) under 10 words to 50.7\% at 10--20 words and 55.1\% (76.9\% CAcc) at 40+ words (+12.7 points), making continuous speech more vulnerable than short commands.

At the sentence level, open-class \emph{content words} (nouns, verbs, adjectives, and adverbs) carry the primary semantic payload of an utterance compared with closed-class \emph{function words} (e.g., articles and prepositions). Under raw CTC decoding at 100\% volume (Figure~\ref{fig:metamic_error_breakdown}, right axis), short function words achieve 68.1\% exact-match recall, whereas longer multi-syllable content words reach 39.9\% recall (34.1\% in hand, 24.8\% at 80\% volume, and 12.6\% at 60\% volume) because even a single character error marks an entire word wrong under exact matching, and longer words offer more positions for character substitutions. Crucially, because raw transcripts still preserve the phonetic skeletons of these words (Table~\ref{tab:wacc_examples}), text-only LLM refinement (Appendix~\ref{append:llm}) uses sentence context to resolve local character substitutions, retaining 92.7\% of raw-correct content words and restoring 27.0\% of raw-incorrect ones. This step reduces $\mathrm{Sub.}$ from 41.4\% to 30.7\%, boosts content-word recall by 13.3 percentage points to 53.2\% (+33.3\% relative; function words rise to 72.4\%), and raises corpus mean WAcc from 52.1\% to 60.1\% ($\mathrm{Del.}=6.1\%$, $\mathrm{Ins.}=3.1\%$; $\mathrm{CAcc}=73.0\%$, as whole-word rewrites alter multiple characters).

Across all 816 test utterances per phone at 100\% volume, the complementary CDFs in Figure~\ref{fig:metamic_accuracy_cdf} show that under raw CTC decoding, 72.6\% of utterances (min--max: 60.2--87.6\%) achieve $\mathrm{WAcc}\ge 40\%$, 56.5\% achieve $\ge 50\%$ (median 50.5\%), and 16.4\% reach $\ge 70\%$. LLM refinement shifts the curve upward across the entire range, raising median WAcc to 60.1\%, the share with $\mathrm{WAcc}\ge 50\%$ to 66.4\%, and the share with $\mathrm{WAcc}\ge 70\%$ to 33.6\%.

\subsubsection{Scaling Linguistic Supervision in $p_S$}
\label{sec:eval_source_scale}

Finally, with $p_\Theta$ fixed, we evaluate how accuracy scales with the source-speech distribution $p_S$ (Equation~\eqref{eq:s2l-objective}) by training GR\&MS on nested 20\%, 40\%, 60\%, 80\%, and 100\% subsets of LibriSpeech \textit{train-960} (Figure~\ref{fig:metamic_training_scale}). Six-phone mean WAcc rises monotonically from 29.6\% to 38.2\%, 44.1\%, 47.7\%, and 52.1\% (CAcc from 60.8\% to 74.3\%; min--max WAcc from 25.3--34.1\% at 20\% to 45.0--61.8\% at 100\%). Notably, just 20\% of this public speech corpus (29.6\% WAcc, zero hardware traces) already exceeds 24-hour profiled ISpyU cross-device accuracy (25.9\%), and 40\% (38.2\%) surpasses ISpyU's same-device mean (36.6\%).
\section{Discussion}
\label{sec:discussion}

\subsection{Countermeasures and Cross-API Governance}
\label{sec:countermeasures}
\noindent\textbf{Aligning interface permissions with exported physical observations.}
While \texttt{CAMERA} authorizes visual capture, exporting high-rate timestamped gyroscope arrays via \texttt{CaptureResult} turns a visual API into an unmediated inertial side channel. Platform permission models should therefore audit secondary metadata by physical sensing fidelity (sampling rate, timestamp precision, and numerical resolution), independent of an API's primary media type.

\noindent\textbf{Unified mediation across HAL and framework boundaries.}
Because Camera2 and \texttt{SensorManager} tap the same upstream MEMS sequence (Finding~2), throttling \texttt{SensorManager} alone leaves Camera2 unmediated (Findings~1 and~4) and lets hardware rate elevation spill across concurrent listeners (Finding~5). Moreover, because OEM gyro tags reside in opaque \texttt{VENDOR\_SECTION} namespaces (often proprietary byte structs), \texttt{CameraMetadataNative} cannot generically parse and downsample them. Closing this gap requires a two-layer fix: (i)~\texttt{CameraService} should strip vendor tags by default for third-party apps (enforced via Android CTS), keeping high-rate stabilization arrays inside the Camera HAL; and (ii)~any standardized user-space motion metadata (e.g., \texttt{STATISTICS\_OIS\_SAMPLES}) must hook into \texttt{SensorPrivacyManager} to enforce the $\le 200$~Hz cap, \texttt{HIGH\_SAMPLING\_RATE\_SENSORS} checks, per-client rate isolation, and privacy toggles.

\noindent\textbf{Effect of sampling-rate tiers and 200~Hz capping.}
To isolate sampling-rate effects, we decimate the six phones' measured Camera2 traces by $2\times$ (400--499~Hz, matching privileged \texttt{SensorManager} and the 20 medium-rate models in Section~\ref{sec:exposure_landscape}) and by $5\times$ (160--200~Hz, within Android's default cap), training a rate-matched S2L recognizer per tier. Mean raw WAcc drops from 52.1\% at 800--1000~Hz to 32.4\% at 400--499~Hz (showing 400--500~Hz exposure still poses moderate risk) and plunges to 2.3\% at 160--200~Hz, confirming that enforcing Android's 200~Hz cap closes the continuous-speech channel.

\subsection{Architectural Scope and Broader Applicability}
\noindent\textbf{Mobile OS architecture (Android vs.\ iOS).}
Unlike Android's \texttt{CameraMetadataNative} forwarding populated vendor HAL tags to user space, iOS's \texttt{AVFoundation} confines stabilization gyroscope streams inside the system media daemon without exposing raw HAL arrays to apps, while \texttt{CMMotionManager} caps public gyroscope delivery at 100~Hz ($10\times$ below 1000~Hz Camera2 streams). Restricting vendor stabilization tags to HAL-internal consumers aligns Android's camera pipeline with this least-privilege boundary.

\noindent\textbf{Physical scope and broader applicability.}
Physically, \sysname targets structure-borne loudspeaker vibrations during stationary or lightly held camera sessions (e.g., video calls or online lectures); earphone audio, unamplified talkers, and ambient room noise impart negligible chassis energy across the air--chassis impedance mismatch~\cite{anand2018speechless,anand2021spearphone}, while sustained rapid camera panning can introduce motion artifacts beyond our low-frequency filter. Beyond acoustic gyroscopes, S2L extends to other side channels with parameterizable forward physics (e.g., rolling-shutter optical, mmWave, and EM analysis), eliminating target profiling across hardware.

\section{Conclusion}
\label{sec:conclusion}

We identify an acoustic side channel in Camera2 \texttt{CaptureResult} gyroscope metadata that bypasses Android's sensor-rate caps and system-wide privacy controls. Across 50 phones, 37 expose Camera2 gyro streams and 26 exceed the 200~Hz cap, including six 800--1000~Hz models that tap the same upstream MEMS sequence as \texttt{SensorManager} yet stream unthrottled without \texttt{HIGH\_SAMPLING\_RATE\_SENSORS} even when microphone and camera privacy toggles are disabled.

To enable profiling-free cross-device speech recognition, we introduce Source-to-Leakage Learning (S2L) in \sysname. Trained solely on physics-simulated leakage, a single \sysname model achieves 52.1\% mean WAcc (up to 61.8\%) and 74.3\% mean CAcc (up to 80.7\%) across the six high-rate phones, improving cross-device WAcc by 101.4\% over fine-tuned ISpyU models (25.9\%) and surpassing 24-hour same-device fine-tuning on every phone. These results show that permission boundaries must mediate all physical observations exported across APIs.

\clearpage
\bibliographystyle{IEEEtran}
\bibliography{references}

\appendices

\section{Camera2 Channel-Family and Implementation Details}
\label{append:model}

\subsection{Channel-Family Parameterization and Sampling}
\label{append:response}
\label{append:sampling-readout}

Table~\ref{tab:camera2-response-prior} summarizes the fixed parameter supports of $p_\Theta$ used to instantiate Equations~\eqref{eq:s2l-composite}--\eqref{eq:ms-readout}. Four frequency-stratified structural modes ($M=4$) span the 100--1000~Hz mechanical resonance band reported in camera and OIS studies~\cite{bk_modal_testing,la2015optical,chiou2010design}. Each pole pair $p_{m,\theta} = -\sigma_{m,\theta} + j\omega_{d,m,\theta}$ in Equation~\eqref{eq:gr-poles} is parameterized by an undamped natural frequency $f_{m,\theta} = \omega_{m,\theta}/(2\pi)$ and an underdamped ratio $\zeta_{m,\theta} \in [0.03, 0.40]$, setting $\sigma_{m,\theta} = \zeta_{m,\theta}\omega_{m,\theta}$ and $\omega_{d,m,\theta} = \omega_{m,\theta}\sqrt{1-\zeta_{m,\theta}^2}$. The 300--800~Hz low-pass cutoff $f_{c,\theta}=\omega_{c,\theta}/(2\pi)$ reflects commercial gyroscope signal paths~\cite{bosch_bmi270,st_lsm6dsox}, and $H_\theta(\omega)$ in Equation~\eqref{eq:gr-operator} is evaluated over the 0--1~kHz band on the $F_w=16$~kHz grid. For irregular metadata readout (Equation~\eqref{eq:ms-readout}), each trace draws an integer lattice stride $h_\theta \in \{15,\dots,21\}$ (bracketing 800--1000~Hz streams with a $\pm 1$-sample margin) and continuous offset $u_\theta\sim\mathcal U(0,0.50)$ to define the nominal period $1/F_\theta = (h_\theta - u_\theta)/F_w$ (762--1103~Hz). Per-sample timing jitter $\delta\tau_i \sim \mathcal N(0,\sigma_{\tau,\theta}^2)$ uses $\sigma_{\tau,\theta}\sim\mathcal U(18.75,43.75)\,\mu\text{s}$ ($0.30\text{--}0.70$ samples at $F_w = 16$~kHz) before adding readout noise and scalar quantization.
\begin{table}[t!]
\centering
\scriptsize
\setlength{\tabcolsep}{3pt}
\renewcommand{\arraystretch}{0.98}
\caption{Fixed parameter support of the Camera2 prior $p_\Theta$ (frequencies in Hz; waveform grid $F_w = 16$~kHz).}
\label{tab:camera2-response-prior}
\begin{tabularx}{\columnwidth}{@{}p{0.27\columnwidth}p{0.34\columnwidth}X@{}}
\toprule
\textbf{Parameter} & \textbf{Role} & \textbf{Support / Setting} \\
\midrule
$M$; $f_1,f_2,f_3,f_4$ & Stratified modal freqs. & $M=4$; $[100,300]$, $[300,500]$, $[500,750]$, $[750,1000]$ \\
$\zeta_{m,\theta}$ & Pole damping ratio & $[0.03, 0.40]$ (yields $p_{m,\theta}$) \\
$|a^{q}_{m,\theta}|$, $q\in\{\mathrm p,\mathrm r\}$ & Modal residues & $[0.10, 4.0]$; $\operatorname{sgn}(a^q_{1,\theta})=+1$, others $\{-1,+1\}$ \\
$\lambda_{\theta}$; $\nu_{\theta}$; $f_{\mathrm{ref}}$ & Readout mix \& deriv.\ order & $\lambda_\theta \in [0.10, 0.70]$; $\nu_\theta \in \{0,1,2\}$; $f_{\mathrm{ref}} = 1000$ \\
$f_{c,\theta}$; $p_{\theta}$ & Roll-off scale \& order & $f_{c,\theta} \in [300, 800]$; $p_\theta \in [3, 10]$ \\
\midrule
$h_\theta$; $u_\theta$ & Nominal period params. & $h_\theta \in \{15,\dots,21\}$; $u_\theta \sim \mathcal U(0, 0.50)$ ($762\text{--}1103$~Hz) \\
$\sigma_{\tau,\theta}$ & Inter-sample timing jitter & $\mathcal U(18.75, 43.75)\,\mu\text{s}$ ($0.30\text{--}0.70$ grid samples) \\
$\sigma_{\eta,\theta}$; $\delta_\theta$ & Noise SNR \& quant.\ step & $\mathrm{SNR}\in [10, 36]$~dB; $\delta_\theta \in [2.5\%, 12.5\%]$ of peak \\
\bottomrule
\end{tabularx}
\end{table}

\begin{table*}[!t]
\centering
\footnotesize
\setlength{\tabcolsep}{3.0pt}
\renewcommand{\arraystretch}{1.0}

\caption{Representative raw X300s predictions across 10-point utterance-level WAcc intervals at 100\% media volume.}
\label{tab:wacc_examples}

\begin{tabularx}{\textwidth}{
@{}
>{\ttfamily\raggedright\arraybackslash}p{0.13\textwidth}
>{\centering\arraybackslash}p{0.06\textwidth}
>{\raggedright\arraybackslash}X
>{\raggedright\arraybackslash}X
@{}}
\toprule
\multicolumn{1}{c}{\textbf{Utterance ID}} &
\textbf{WAcc (\%)} &
\multicolumn{1}{c}{\textbf{Reference Transcript}} &
\multicolumn{1}{c}{\textbf{Raw Prediction}} \\
\midrule

1462-170138-0009
& 0.0
& its delightful to hear it in a london theatre
& this delit was be her uti an e dae \\

422-122949-0002
& 10.0
& we truthful ones the nobility in ancient greece called themselves
& we tuthfor mon fon livitnacence to en sa \\

6241-61946-0012
& 20.0
& a few stray cows and sheep were only seen occasionally
& a few stare casand peper only to n ul on \\

6241-61943-0019
& 30.0
& in the meantime there is not an hour to lose
& in the men time ire is on hi to leve \\

2035-147961-0010
& 40.0
& he seemed to be cursing people who had wronged him
& he seemd be courly people wold wise him \\

251-137823-0000
& 50.0
& ill be glad to try sir he replied
& i ll be glad to sty sir she replied \\

3853-163249-0040
& 60.0
& now lets be brave and enjoy every minute of it
& sew lef te bols and enjoy every minute of it \\

2277-149897-0015
& 70.0
& in about an hour and three quarters the boy returned
& ind about an hour and three foathers the boy wasturne \\

1462-170145-0014
& 80.0
& but when i came i thought i had been mistaken
& but when i tame i thought i had ben mistaken \\

2277-149896-0012
& 90.0
& she had not been able to get away this morning
& she had not been able to get away this sendement \\

2428-83705-0000
& 100.0
& her father is a most remarkable person to say the least
& her father is a most remarkable person to say the least \\

\bottomrule
\end{tabularx}
\end{table*}    

\subsection{Representation, Architecture, and Training}
\label{append:representation}
\label{append:recognizer}
\label{append:implement}

In Equation~\eqref{eq:representation-norm}, each centered axis $q\in\{x,y,z\}$ is converted into a linear-frequency STFT magnitude spectrogram $S_q(f,t)$, discarding the lowest frequency bins to form $\mathcal B$ and suppress low-frequency motion artifacts. The acoustic recognizer adopts a standard CNN--BiLSTM--CTC architecture comprising convolutional front-end blocks, stacked bidirectional LSTM layers, and a fully connected output projection with CTC loss. We train the recognizer on the LibriSpeech \textit{train-960} corpus with SpecAugment~\cite{park2019specaugment} using standard AdamW optimization and greedy CTC decoding.

\section{Measurement and Evaluation Details}
\label{sec:appendix_measurement}

\subsection{Qualitative WAcc Examples and Text Correction}
\label{append:wacc_examples}
\label{append:llm}

Table~\ref{tab:wacc_examples} provides representative raw X300s predictions at 10-percentage-point WAcc intervals under 100\% media volume. The reference transcripts contain 8--11 words, and all text retains the lowercase, punctuation-free normalization used for scoring.

The 50\% example, close to the corpus-level mean WAcc, illustrates the relationship between edit-distance accuracy and readability. Five of the eight reference words match exactly (62.5\%), yet WAcc is 50\% because word-level edit distance penalizes insertions alongside substitutions: the alignment contains three substitutions (\textit{ill}$\rightarrow$\textit{i}, \textit{try}$\rightarrow$\textit{sty}, and \textit{he}$\rightarrow$\textit{she}) and one insertion (\textit{ll}), yielding four edits over eight reference words. The same prediction reaches 87.9\% CAcc, illustrating that word-level errors can coexist with substantial character-level preservation.

For text-only LLM refinement, \texttt{gpt-5.4} receives only the raw CTC-decoded transcript with a fixed instruction to produce a coherent, punctuation-free English sentence without extra commentary. We request one completion per utterance without candidate selection or manual editing, retaining the raw prediction if the returned completion is empty.

\subsection{Device Matrix, Rate Estimation, and Protocols}
\label{sec:appendix_device_matrix}
\label{sec:appendix_rate_estimation}
\label{append:fidelity_variants}
\label{append:fidelity-protocol}
\label{append:ispyu-reproduction}

\noindent\textbf{Device matrix and rate estimation.}
Table~\ref{tab:appendix_device_matrix} reports the Camera2 vendor metadata exposure across all 50 tested phones. For timestamped streams, after restoring chronological order and removing duplicate timestamps, we convert embedded timestamps $t_i$ to seconds and estimate the effective sampling rate as $\widehat F = (\operatorname{median}_{i}\{t_i-t_{i-1}:t_i>t_{i-1}\})^{-1}$, cross-checked against valid sample counts and frame cadence (reporting the $\approx 30$~Hz callback rate for frame-bound fields).

\noindent\textbf{Prior-work reproduction (ISpyU).}
Following ISpyU's two-stage synthesis-and-adaptation procedure~\cite{zhang2023spy}, we generate 281{,}241 synthetic training samples from LibriSpeech \textit{train-960} via uniform $16\times$ decimation (matching the 1-kHz Camera2 band, analogous to ISpyU's $32\times$ factor at 500~Hz) to pretrain the 50-epoch ISpyU-sim base checkpoint with SpecAugment~\cite{park2019specaugment}. For each single-device ISpyU baseline in Table~\ref{tab:ispyu_asr}, we fine-tune this checkpoint on 7{,}681 real Camera2 traces ($\approx 24$~hours from LibriSpeech \textit{train-960}, speaker-disjoint from all 16 \textit{dev-clean} test speakers) collected on one profiling phone $d_i$. For the five-phone leave-one-out baseline in Table~\ref{tab:ispyu_lodo}, we partition the same 24-hour adaptation set into disjoint per-phone subsets across the five non-target phones and test on the held-out sixth phone. All replay-recorded traces follow Section~\ref{sec:eval_setup} and are evaluated on the held-out 816 real traces per phone.

\noindent\textbf{RQ2 ablation and spectrogram-fidelity protocols.}
In RQ2, all four simulator variants (ISpyU-sim, GR, MS, and GR\&MS) train the identical CNN--BiLSTM--CTC recognizer for 50 epochs under the same optimizer and SpecAugment schedule; greedy CTC decoding maps frame-level spectrogram features directly to character sequences, tying downstream accuracy directly to the physical speech structure preserved by each simulator. For the spectrogram-fidelity diagnostic on 50 randomly sampled held-out utterances per phone, we hold the selected measured axis fixed across variants, set the nominal sampling rate of MS and GR\&MS to each trace's estimated rate $\widehat F$, and compute the peak 2D Pearson correlation coefficient $r_{\mathrm{spec}}$ between each normalized measured STFT spectrogram $\widetilde S(f,t)$ and $K=32$ draws from $p_\Theta$ (using a $1\times10^{-3}$ floor for zero-variance frames).

\begin{table}[!ht]
\centering
\footnotesize
\setlength{\tabcolsep}{3.8pt}
\renewcommand{\arraystretch}{1.0}
\caption{Five-phone leave-one-out ISpyU pooling (24-hour total budget) versus zero-profiling \sysname.}
\label{tab:ispyu_lodo}
\begin{tabular}{@{} l c c c c @{}}
\toprule
\textbf{Handset} & \multicolumn{2}{c}{\textbf{Five-Phone ISpyU~\cite{zhang2023spy}}} & \multicolumn{2}{c}{\textbf{\sysname (Ours)}} \\
\cmidrule(lr){2-3}\cmidrule(l){4-5}
 & \textbf{WAcc (\%)} & \textbf{CAcc (\%)} & \textbf{WAcc (\%)} & \textbf{CAcc (\%)} \\
\midrule
X300s & 47.4 & 67.1 & \textbf{61.8} & \textbf{80.7} \\
X300P & 31.3 & 55.1 & \textbf{45.0} & \textbf{69.3} \\
X200U & 23.5 & 50.2 & \textbf{50.6} & \textbf{73.6} \\
17TP  & 26.2 & 51.7 & \textbf{51.1} & \textbf{73.1} \\
M8PA  & 24.6 & 50.4 & \textbf{48.9} & \textbf{72.6} \\
K90M  & 25.0 & 50.9 & \textbf{55.3} & \textbf{76.2} \\
\midrule
\textbf{Mean} & 29.7 & 54.2 & \textbf{52.1} & \textbf{74.3} \\
\bottomrule
\end{tabular}
\end{table}

\newcolumntype{W}[1]{%
  >{\hsize=#1\hsize\linewidth=\hsize\raggedright\arraybackslash}X}
\newcolumntype{Y}[1]{%
  >{\hsize=#1\hsize\linewidth=\hsize\centering\arraybackslash}X}

\begin{table*}[t]
\centering
\footnotesize
\setlength{\tabcolsep}{1.5pt}
\renewcommand{\arraystretch}{0.95}

\caption{Gyro-related Camera2 vendor metadata across the 50 tested phones. OIS-only metadata is omitted; a dash indicates that no usable stream was recovered.}
\label{tab:appendix_device_matrix}

\begin{tabularx}{\textwidth}{
@{}
W{1.05}
W{1.35}
Y{0.45}
W{1.55}
Y{0.55}
W{1.05}
@{}}
\toprule
\textbf{OEM Group} &
\textbf{Model} &
\textbf{Year} &
\textbf{SoC Family} &
\textbf{Rate} &
\textbf{Key(s)} \\
\midrule

\textbf{Google/Pixel}
& Google Pixel 6a
& 2022
& Google Tensor
& --
& -- \\

& Google Pixel 9
& 2024
& Google Tensor G4
& --
& -- \\

\midrule

\textbf{Xiaomi/Redmi}
& Xiaomi 13 Ultra
& 2023
& Snapdragon 8 Gen 2
& --
& -- \\

& Xiaomi 14
& 2023
& Snapdragon 8 Gen 3
& --
& -- \\

& Xiaomi 14 Pro
& 2023
& Snapdragon 8 Gen 3
& 460~Hz
& XMI \\

& Redmi K80 Pro
& 2024
& Snapdragon 8 Elite
& 460~Hz
& XMI \\

& Redmi K90 Pro Max
& 2025
& Snapdragon 8 Elite Gen 5
& 30~Hz
& XMI-TP \\

& Redmi Turbo 5 Max
& 2026
& Dimensity 9500s
& 500~Hz
& MTK \\

& \textbf{Redmi K90 Max}
& 2026
& Dimensity 9500
& \textbf{800~Hz}
& MTK \\

& \textbf{Xiaomi 17T Pro}
& 2026
& Dimensity 9500
& \textbf{800~Hz}
& MTK \\

\midrule

\textbf{Samsung}
& Galaxy S23 Ultra
& 2023
& Snapdragon 8 Gen 2 for Galaxy
& 30~Hz
& SAM \\

& Galaxy S24 Ultra
& 2024
& Snapdragon 8 Gen 3 for Galaxy
& 30~Hz
& SAM \\

& Galaxy A55 5G
& 2024
& Exynos 1480
& --
& -- \\

& Galaxy S25
& 2025
& Snapdragon 8 Elite for Galaxy
& 30~Hz
& SAM \\

\midrule

\textbf{OPPO/OnePlus/realme}
& OPPO Find X7 Ultra
& 2024
& Snapdragon 8 Gen 3
& 30~Hz
& OPL \\

& OPPO Find X8
& 2024
& Dimensity 9400
& 400~Hz
& MTK \\

& OPPO Find N5
& 2025
& Snapdragon 8 Elite
& 30~Hz
& OPL \\

& OPPO Find X8S
& 2025
& Dimensity 9400+
& 400~Hz
& MTK \\

& OPPO Find X9
& 2025
& Dimensity 9500
& 400~Hz
& MTK \\

& OPPO Find X9 Pro
& 2025
& Dimensity 9500
& 400~Hz
& MTK \\

& OPPO K15 Pro
& 2026
& Dimensity 8500 SUPER
& 400~Hz
& MTK \\

& realme GT 8 Pro
& 2025
& Snapdragon 8 Elite Gen 5
& 30~Hz
& OPL \\

& OnePlus Ace 6T
& 2025
& Snapdragon 8 Gen 5
& 30~Hz
& OPL \\

& OnePlus Ace 6 Ultra
& 2026
& Dimensity 9500
& 500~Hz
& MTK \\

\midrule

\textbf{vivo/iQOO}
& vivo X100
& 2023
& Dimensity 9300
& 500~Hz
& VIVO-CG \\

& vivo X100 Pro
& 2023
& Dimensity 9300
& 500~Hz
& VIVO-CG \\

& vivo S18
& 2023
& Snapdragon 7 Gen 3
& 100~Hz
& VIVO-G \\

& iQOO Neo9 Pro
& 2023
& Dimensity 9300
& 400~Hz
& MTK; VIVO-CG; VIVO-G \\

& vivo X Fold3 Pro
& 2024
& Snapdragon 8 Gen 3
& 100~Hz
& VIVO-G \\

& vivo X100s
& 2024
& Dimensity 9300+
& 500~Hz
& VIVO-CG \\

& vivo X100s Pro
& 2024
& Dimensity 9300+
& 500~Hz
& VIVO-CG \\

& iQOO Neo9S Pro
& 2024
& Dimensity 9300+
& 400~Hz
& MTK \\

& iQOO Neo9S Pro+
& 2024
& Snapdragon 8 Gen 3
& 500~Hz
& VIVO-CG \\

& \textbf{vivo X200 Ultra}
& 2025
& Snapdragon 8 Elite
& \textbf{910~Hz}
& VIVO-CG \\

& iQOO Neo 10 Pro+
& 2025
& Snapdragon 8 Elite
& 90~Hz
& VIVO-CG; VIVO-G \\

& vivo S30 Pro mini
& 2025
& Dimensity 9300+
& 500~Hz
& VIVO-CG \\

& vivo X Fold5
& 2025
& Snapdragon 8 Gen 3
& 500~Hz
& VIVO-CG \\

& \textbf{vivo X300 Pro}
& 2025
& Dimensity 9500
& \textbf{960~Hz}
& MTK; VIVO-CG \\

& iQOO 15
& 2025
& Snapdragon 8 Elite Gen 5
& --
& -- \\

& iQOO Neo 11
& 2025
& Snapdragon 8 Elite
& 500~Hz
& VIVO-CG \\

& \textbf{vivo X300s}
& 2026
& Dimensity 9500
& \textbf{1000~Hz}
& MTK \\

& vivo X300 Ultra
& 2026
& Snapdragon 8 Elite Gen 5
& --
& -- \\

& iQOO 15T
& 2026
& Dimensity 9500 Monster
& 500~Hz
& MTK; VIVO-CG \\

\midrule

\textbf{Huawei}
& HUAWEI P40
& 2020
& Kirin 990 5G
& --
& -- \\

\midrule

\textbf{HONOR}
& HONOR X40
& 2022
& Snapdragon 695
& --
& -- \\

& HONOR 400
& 2025
& Snapdragon 7 Gen 4
& --
& -- \\

& \textbf{HONOR Magic8 Pro Air}
& 2026
& Dimensity 9500
& \textbf{800~Hz}
& MTK \\

\midrule

\textbf{Motorola/Lenovo}
& Motorola moto X70 Air Pro
& 2026
& Snapdragon 8 Gen 5
& --
& -- \\

\midrule

\textbf{Nubia/ZTE}
& Nubia Z70 Ultra
& 2024
& Snapdragon 8 Elite
& --
& -- \\

\midrule

\textbf{ROG/ASUS}
& ROG Phone 8 Pro
& 2024
& Snapdragon 8 Gen 3
& --
& -- \\

\bottomrule
\end{tabularx}

\vspace{0.3ex}
\begin{minipage}{0.98\textwidth}
\scriptsize
\textit{Key aliases.}
MTK = \texttt{com.mediatek.3afeature.gyrodata};
XMI = \texttt{xiaomi.mivi.ellc.gyro};
XMI-TP = \texttt{com.xiaomi.ThridPartyGyroState.gyroData};
SAM = \texttt{samsung.android.listener.gyro.gyroValues};
OPL = \texttt{com.oplus.gyro.data};
VIVO-CG = \texttt{vivo.parameter.captureGyroData};
VIVO-G = \texttt{vivo.parameter.gyroData}.
OIS-only keys, including \texttt{vivo.parameter.captureOisData}, are omitted.
\textbf{Bold nominal rates} indicate the six phones used for continuous-speech evaluation.
\end{minipage}

\end{table*}

\end{document}